\documentclass[secnumarabic, twocolumn, graphics,floatfix, superscriptaddress,tightenlines, aps, pra, longbibliography]{revtex4-1}

\usepackage{bm}
\usepackage{epsfig}
\usepackage{float}
\usepackage{color}
\usepackage[english]{babel}
\usepackage[]{natbib} 
\usepackage{lipsum}
\usepackage{amsmath}
\usepackage{listings}
\usepackage{booktabs}
\usepackage{textcomp}

\begin{document}

\title[Complexity of dipolar exciton  Mott  transition  in GaN/(AlGa)N nanostructures  ]
  {Complexity of dipolar exciton  Mott  transition  in GaN/(AlGa)N nanostructures 
}

\author{F. Chiaruttini}
\affiliation{Laboratoire Charles Coulomb, UMR 5221
 CNRS-Universit\'{e} de Montpellier, 34095 Montpellier, France}
 \author{T. ~Guillet}
\affiliation{Laboratoire Charles Coulomb, UMR 5221
 CNRS-Universit\'{e} de Montpellier, 34095 Montpellier, France}
 \author{C.~Brimont}
\affiliation{Laboratoire Charles Coulomb, UMR 5221
 CNRS-Universit\'{e} de Montpellier, 34095 Montpellier, France}
  \author{D.~Scalbert}
\affiliation{Laboratoire Charles Coulomb, UMR 5221
 CNRS-Universit\'{e} de Montpellier, 34095 Montpellier, France}
  \author{S.~Cronenberger}
\affiliation{Laboratoire Charles Coulomb, UMR 5221
 CNRS-Universit\'{e} de Montpellier, 34095 Montpellier, France}
  \author{B.~Jouault}
\affiliation{Laboratoire Charles Coulomb, UMR 5221
 CNRS-Universit\'{e} de Montpellier, 34095 Montpellier, France}\author{P.~Lefebvre}
\affiliation{Laboratoire Charles Coulomb, UMR 5221
 CNRS-Universit\'{e} de Montpellier, 34095 Montpellier, France}
\author{B. Damilano}
\affiliation
{CRHEA, Universit\'e C\^ote d’Azur, CNRS, Valbonne, France}
 \author{M.~Vladimirova}
\affiliation{Laboratoire Charles Coulomb, UMR 5221
 CNRS-Universit\'{e} de Montpellier, 34095 Montpellier, France}



%
%
%
%
%

\begin{abstract}%
The  Mott transition from a dipolar excitonic liquid to an electron-hole plasma is demonstrated in a wide GaN/(Al,Ga)N quantum well at $T=7$~K by means of spatially-resolved magneto-photoluminescence spectroscopy. 
Increasing optical excitation density we drive the system from the excitonic state, characterized by a diamagnetic behavior and thus a quadratic energy dependence on the magnetic field, to the unbound electron-hole state, characterized by a linear shift of the emission energy with the magnetic field.
The complexity of the system requires to take into account both the density-dependence of the exciton binding energy and the exciton-exciton interaction and correlation energy that are of the same order of magnitude.  We estimate the carrier density at Mott transition  as $n_\mathrm{Mott}\approx 2\times 10^{11}$~cm$^{-2}$ and address  the role played by excitonic correlations in this process. 
Our results strongly rely on the spatial resolution of the photoluminescence and the assessment of the carrier transport. We show, that in contrast to GaAs/(Al,Ga)As systems, where transport of dipolar magnetoexcitons is strongly quenched by the magnetic field due to exciton mass enhancement, in GaN/(Al,Ga)N the band parameters are such that the transport is preserved up to $9$~T.

\end{abstract}
\maketitle
\section{Introduction}
\begin{figure} [H]
\includegraphics [width=1\columnwidth] {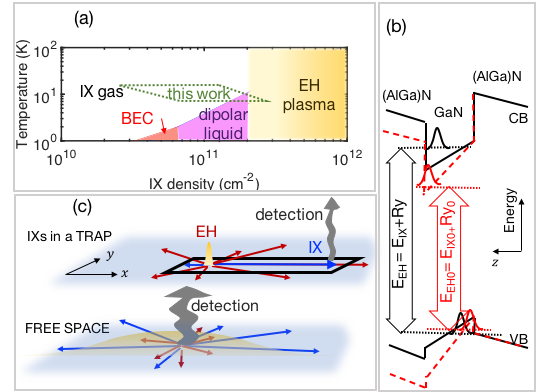}
  \caption{ (a) IX phase diagram. Four phases theoretically expected are shown by different colors.  The green region delimited by dotted line represents the phase space area explored in this work. The critical density corresponding to the transition towards plasma states is determined in this work. 
  (b) Sketch of the the GaN/(Al,Ga)N QW  bands in the absence (red dashed line)  and in the presence of  optical excitation (black solid line). 
  Corresponding electron and hole wavefunctions and energy levels are shown with the same color code. 
The band-to-band  transitions are indicated by arrows.
  (c) Artist view of the two ways to study IX emission:  "IX in a trap" (upper sketch) - point-like excitation followed by spatially separated PL analysis in the trap (used in \cite{Chiaruttini2019}) and "free space"  (lower sketch) - broad excitation followed by detection in the center of the excitation spot (used in this work). }
  \label{fig:fig1} 
\end{figure}

New states of matter and various phase transitions in bosonic systems have benefitted from particular attention in recent years. 
Among different physical platforms, two-dimensional bilayer systems, where electron and hole  are spatially separated and form an indirect exciton (IX), are particularly interesting.
Due to this spatial separation between electron and hole, the lifetime of the IX is considerably increased (up to tens of microseconds),  and it acquires a non-zero electric dipole moment \cite{LozovikYudson,Miller1985,Huber1998,Ivanov1999,Butov1999,High2008,Butov2017}.
These rich and tuneable properties offer the opportunity to achieve experimentally high densities of cold IXs, and to explore various intriguing quantum phenomena including   Bose-Einstein-like condensation (BEC), darkening, formation of dipolar liquids, superfluidity and excitonic ferromagnetism \cite{Lozovik1996,High2012,Shilo2013,Schinner2013,Cohen2016,Butov2016,Combescot2017,Anankine2017,Misra2018}. 
A possible phase diagram that can be suggested theoretically for IXs hosted by GaN/(AlGa)N quantum wells (QWs) is represented in Fig.~\ref{fig:fig1}~(a)~\cite{Laikhtman2009}.

 Fig.~\ref{fig:fig1}~(a) shows  that the  critical temperature that should be attained to scrutinise collective IX states is constrained by the maximum possible IX density. 
The difficulty faced in this context is the onset of avalanche ionization of excitons at high densities, as a result of screening and momentum space filling~\cite{Zimmermann1978}.
This many-body effect is usually referred to as the Mott transition, in analogy with the phase transition from an electrically insulating to a conducting state of matter predicted by Sir Nevill Mott in a system of correlated electrons \cite{Mott1968}.
Mott transition of IXs has been addressed  both experimentally and theoretically, but mainly in  GaAs-based  coupled quantum wells under electric bias \cite{BenTaboudeLeon2003,Nikolaev2004,Stern2008,Snoke2008,Byrnes2010,Kirsanske2016,Vignesh2020}. 
However, so far, no consensus regarding the dynamics of this transition in  two-dimensional systems has been reached, neither for traditional excitons, nor  for IXs. 
In particular, it is not clear whether the ionization  of excitons occurs abruptly or gradually when the density is increased \cite{Nikolaev2004,Kappei2005,Deveaud2005,Nikolaev2008,Stern2008,Snoke2008,Sekiguchi2017,Vignesh2020}. 
Moreover, an intriguing  hysteretic behavior has been predicted when the temperature is slowly changed at an intermediate carrier density \cite{Nikolaev2008}, while at low densities an entropic ionisation  of excitons can be expected \cite{Mock1978}.  

The objective of this work is to address the Mott transition of IXs hosted by wide polar GaN/(AlGa)N quantum wells.
%
%
GaN-hosted  IXs have relatively high binding energies ($20$~meV), small Bohr radii ($5$~nm) and greater mass $M=1.6\,m_0$, as compared to GaAs-based structures ($m_0$ is the free electron mass) \cite{Gil2014}. 
Hence the density/temperature phase diagram (Fig. \ref{fig:fig1}~(a)) is quite different from that of GaAs-based systems. 
%

Because GaN QWs grown along the (0001) crystal axis naturally exhibit built-in electric fields of the order of megavolts per centimeter, IXs are naturally created in the absence of any external electric field \cite{Bernardini1997,Leroux1998,Grandjean1999}. 
This offers the opportunity not only for all-optical generation of IXs,  but also for shaping at will the in-plane potential experienced by the IXs by simple deposition of metallic layers on the surface, without application of any external electric bias.
We have recently  demonstrated the electrostatic trapping and control of the cold IX density  in such structures~\cite{Chiaruttini2019}. 
IXs at quasi-thermal equilibrium state could be studied in spatial regions situated tens of micrometers away from the laser excitation spot. 
We called such areas "excitonic lakes". 
We have determined the lower limit for the carrier density for the set of excitation powers used and demonstrated  that the carrier temperature remains power-independent. 
However, despite careful analysis of the spectral shape of the photoluminescence (PL) from the lakes, we could not evidence the onset of the Mott transition.
%

The reasons for that are numerous. 
First, the PL lineshape does not show any characteristic shape modification when the system undergoes Mott transition \cite{Kash1991}. 
Second, excitonic correlations, when they are strong, strongly affect the exciton density estimated by comparing the PL  energy shift to the one calculated by solving the coupled  Schr\"{o}﻿﻿dinger and Poisson equations  (see Fig. \ref{fig:EintComputed}) \cite{Zimmermann,Laikhtman,Schinner2013,Cohen2016,MazuzHarpaz2017}.
Such correlations can arise from the strong depletion of the exciton gas around a given exciton due to repulsive interactions between IXs.
At high IX density they are also expected to favor the transition from an excitonic gas to a dipolar liquid, rather than a Bose-Einstein-like state, see Fig.~\ref{fig:fig1}~(a).
Third, the IX binding energy at the lowest excitation densities, and the typical blue shift of the electron-hole recombination energy at highest excitation densities are of the same order of magnitude, $\approx 20$~meV. 
This makes it extremely challenging to unravel the effects of the binding energy reduction, due to the screening of the built-in electric field and exciton-exciton interactions at non-zero density. 
The density-induced modifications of the QW band diagram and the resulting transition energy shift $E_{int}$ are schematically shown in Fig.~\ref{fig:fig1}~(b).
Finally, the contribution of the band-to-band emission  (unbound electron-hole pairs) to the PL spectrum, that could help us estimating the IX binding energy at least at the lowest excitation densities \cite{Deveaud2005,Kappei2005,Kirsanske2016} is useless in the trap geometry with point-like excitation which we have employed in \cite{Chiaruttini2019}. 
Indeed, when PL is recorded from the same spatial region where the excitation takes place, according to the Saha equation, a mass-action law describing the equilibrium concentration of a gas of coexisting excitons and free carriers, one can eventually extract both exciton and band-to-band emission \cite{ZimmermannBook}. 
This is not the case for a  point-like excitation of IXs.  
Because IXs are rapidly expelled from the excitation spot by strong dipole-dipole interaction, the equilibrium between IXs and band carriers is probably never achieved under the excitation spot (Fig. \ref{fig:fig1}~(c), upper sketch).
The detection of the thermalized IXs trapped in the lake tens of micrometers away from the excitation spot does not offer the opportunity to detect simultaneously  thermalized IX and band-to-band emission.

In this work we choose a different strategy, based on two key ingredients:  (i) broad ($\approx 300$~$\mu$m) and spatially homogeneous excitation, PL detection with spatial resolution, in order to selectively address  the emission in the center of the excitation area (Fig. \ref{fig:fig1}~(c), lower sketch). 
A similar approach has been used in \cite{Kappei2005}. The underlying idea  is to detect the PL of  IXs  and free electron-hole pairs coexisting spatially, and to monitor their relative energies and intensities  as a function of power and temperature;
(ii) application of a magnetic field in Faraday geometry (parallel to the growth axis) in order to discriminate between
the diamagnetic field dependence typical of excitons (quadratic energy shift  { \footnote {Note, that so-called magnetoexcitons characterised by linear energy shift under magnetic field are routinely observed in GaAs when magnetic length becomes  smaller than exciton Bohr radius. This regime is never achieved in this work}}), from the Landau quantization expected for unbound charge carriers (linear increase of energy, usually referred to as cyclotron shift). 

The analysis of the experimental results suggests that at the lowest studied sample temperature ($7$~K) and at the highest excitation power, the carrier density  $n\approx2\times 10^{11}$~cm$^{-2}$ and the system undergoes the Mott transition: the zero-field PL energy spectrally merges with the electron-hole emission and the magnetic field dependence changes frrom quadratic to linear.
Such a high carrier density could not be reached at $T>7$~K, presumably due to more efficient nonradiative recombination.
The determination of the corresponding carrier density  is quite complex, because of the interplay between different density-dependent effects: the built-in electrostatic potential, reduction of the exciton binding energy (accompanied by the increase of the in-plane Bohr radius) and build-up of excitonic correlations. 
In this work, we have tried to unravel their respective contributions. 
We have also observed that, in contrast to GaAs/(AlGa)As systems, where the transport of dipolar magnetoexcitons is strongly quenched by the magnetic field due to exciton mass enhancement \cite{Kuznetsova2017,Dorow2017}, in GaN/(Al,Ga)N the band parameters are such that the transport is preserved up to $9$~T.
%


%
The paper is organised as follows.
In the next Section we present the sample and experimental setup. 
The third Section presents the ensemble of the experimental results, namely the PL study of IX transport as a function of the magnetic field, laser excitation power and temperature. 
In Section  \ref{sec:discussion}  we interpret the results in terms of the situation of our experiments in the IX phase diagram. 
The last section summarises the results and concludes the study.
\begin{figure} 
\includegraphics [width=1\columnwidth] {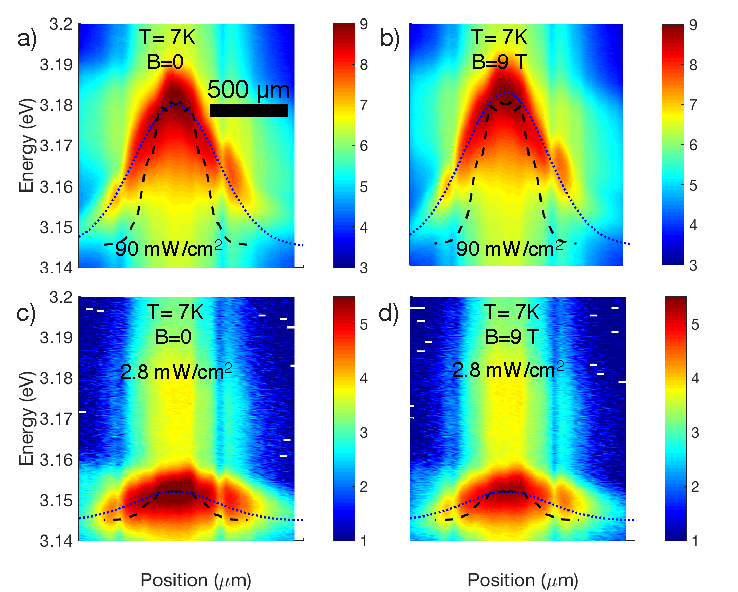}
  \caption{PL intensities (color-encoded in log scale) measured at $T=7$~K and two different excitation power densities: $P=90$~mW/cm$^2$ (a,b),  $P=2.8$~mW/cm$^2$ (c,d), and two magnetic field values: $B=0$ (a,c), $B=9$~T (b,d). 
  The measured laser intensity profile (normlized to unity) is shown by  dashed lines. Dotted lines are Gaussian-shaped guides for the eye showing the decrease of the IX emission energy when the distance from the excitation center increases.}
  \label{fig:transport7K}
\end{figure} 
\begin{figure}
\includegraphics [width=1\columnwidth] {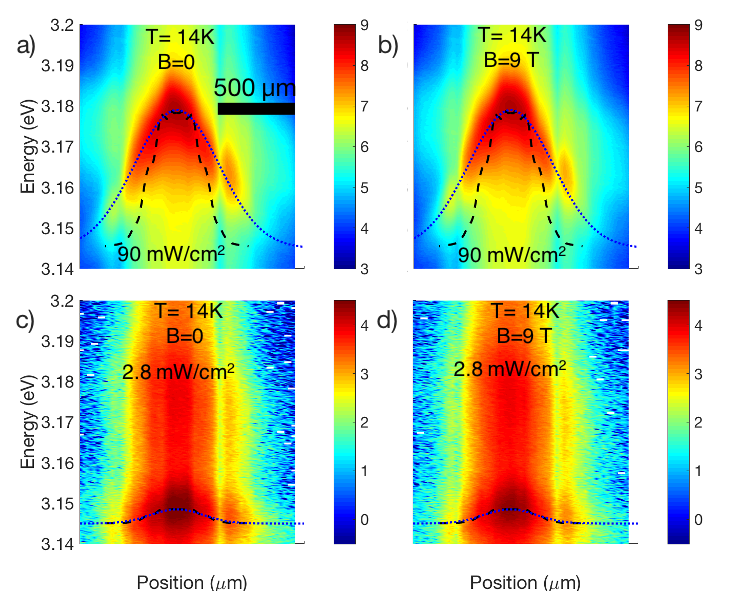}
  \caption{PL intensities (color-encoded in log scale) measured at $T=14$~K and two different excitation power densities: $P=90$~mW/cm$^2$ (a,b),  $P=2.8$~mW/cm$^2$ (c,d), and two magnetic field values: $B=0$ (a,c), $B=9$~T (b,d).  The measured laser intensity profile (normalized to unity) is shown by  dashed lines. Dotted lines are Gaussian-shaped guides for the eye showing the decrease of the IX emission energy when the distance from the excitation center increases.}
  \label{fig:transport14K}
\end{figure} 
\section{Sample and experimental setup}
We investigate a $7.8$~nm-wide GaN QW sandwiched between  
$50$ (top) and $100$~nm-wide (bottom) Al$_{0.11}$Ga$_{0.89}$N barriers, grown on a free-standing GaN substrate.
%
The band diagram of such structure is shown schematically in Fig.\ref{fig:fig1}~(b). 
In the absence of photoexcitation the built-in electric field due to spontaneous and piezoelectric polarization confines electrons and holes at the two opposite interfaces of the QW.  
In the studied structure the field effect is so  strong that the ground state exciton energy ($E_{IX0}=3.145$~eV in Fig.~\ref{fig:fig1}~(b)) is pushed below the exciton energy in bulk GaN $E_{GaN}=3.49$~eV \cite{Bernardini1997,Leroux1998,Grandjean1999,Grandjean2000,Gil2014,Chiaruttini2019}.
The corresponding IX radiative lifetime  is estimated as $\tau_0 \approx 10$~$\mu$s,  orders of magnitude longer than in traditional narrow QWs \cite{Lefebvre1999,Fedichkin2016,Rosales2013,Rossbach2014}.
The sample is placed in a variable temperature magneto-optical cryostat and cooled down to temperatures in the range from $T=7$ to $14$~K. 
Magnetic fields $B$ up to $9$~T created by superconducting magnets are applied in the Faraday geometry (parallel to the QW growth axis). 
 We use a continuous wave laser excitation at $\lambda=355$~nm, very close to the GaN bandgap energy, but well 
 below the emission energy of the Al$_{0.11}$Ga$_{0.89}$N barriers.
 The laser beam is loosely focused on the sample surface ($\approx 300$~$\mu$m-diameter spot) in order to create a broad, homogeneously excited area (see dashed lines in Figs.~\ref{fig:transport7K} and  \ref{fig:transport14K}). 
 The PL spectra are acquired  by a charge coupled device (CCD) camera over $1.5$~mm with $\approx 25$~$\mu$m spatial resolution (pixel size). 
Such broad excitation combined with the spatially resolved detection appeared to be critically important to address the question of the Mott transition. 
 As anticipated in the introduction and as we explain in more details in the next Section, this allows us to detect the PL from IXs and from unbound electron-hole pairs that coexist spatially, and thus to monitor the reduction of the IX binding energy. 
 This also gives us access to carrier transport properties, so that we can directly verify the assumptions regarding transport in our modelling. 
 More details on the sample parameters and on the setup are given in the Appendix.

\section{Experimental results and discussion}
\subsection{Exciton transport}
\label{sec:transport}
The exciton transport at $T=7$~K and $T=14$~K is  presented in Figs.~\ref{fig:transport7K} and \ref{fig:transport14K}, respectively.
They show color-encoded  PL  spectra, measured at various distances from the excitation spot for two different power densities, and  at two values of the magnetic field, $B=0$ and $9$~T. 
Excitation spot intensity profile is shown by the dashed lines. 
The blue dotted lines are guides for the eye, helping to visualise the IX emission peak energy as a function of the distance from the spot center. 

Let us first concentrate on the $B=0$ transport.
Clearly, the energy of the excitonic PL decreases when moving away from the excitation spot center, similarly to the case of the point-like excitation \cite{Fedichkin2015,Fedichkin2016,Chiaruttini2019}.
Such decrease of the emission energy and intensity indicates that the exciton density decreases. 
The relation between PL energy and IX density is particularly strong in wide QWs due to  interactions between IXs that have large dipole moment, and the resulting  screening of the electric field along the growth direction: the highest density under the pumping spot corresponds to the maximum screening  (lowest electric field), and thus to the highest emission intensity and energy \cite{Lefebvre2004,Fedichkin2015,Fedichkin2016,Chiaruttini2019}. 
%
The interaction-induced increase of the IX energy  reaches $40$~meV ($10$~meV) at the spot center and at highest (lowest) power densities that we could explore in our experiments.
This shift can be used to estimate the IX density as  $n=E_{int}/\phi_0$, where 
$\phi_0=1.5\times10^{-13}$~eV$\cdot$cm$^{-2}$ is the coefficient extracted from the self-consistent 
solution of the Schr\"{o}﻿﻿dinger and Poisson equations  (plate capacitor-like model) \cite{Lefebvre2004,Fedichkin2015,Fedichkin2016, Chiaruttini2019}. 
This estimate  needs to be improved by taking into account the exciton binding energy (which decreases when the IX density increases) and  excitonic correlations, which become signicant at low temperatures and high densities. 
As we will discuss later on (see also Figs. \ref{fig:summary}, \ref{fig:EintComputed}),  both of these effects may substantially alter our estimation of the IX density needed to induce a given blue shift ~\cite{Zimmermann,Laikhtman,Laikhtman2009}. 

%

%
%
%
From the $B=0$ emission profiles measured at highest power density $P=90$~mW/cm$^2$ we deduce that despite spatially broad excitation as compared to previous experiments involving point-like excitation, IX transport is still clearly observable both at $7$~K and $14$~K \cite{Fedichkin2015,Fedichkin2016,Chiaruttini2019}. 
The  full width at half maximum (FWHM) of the emission pattern is approximately $800$~$\mu$m.
By contrast, for the lowest power density $P=2.8$~mW/cm$^2$ the IX transport is only apparent at $T=7$~K (same FWHM as at high power), while at $T=14$~K the emission profile coincides with the excitation one. 
The quenching of the IX transport with decreasing power observed at $T=14$~K is a usually observed and well understood phenomenon \cite{Remeika2009,Leonard2012,Rapaport2006,Fedichkin2016}.
Modelling IX transport by drift-diffusion equation shows that this is due to weaker dipole-dipole repulsion, that depends on the density.
More efficient localization on the QW interfaces at low IX densities may also hinder the transport  \cite{Ivanov2002,Remeika2009,Leonard2012,Rapaport2006,Fedichkin2015,Fedichkin2016}.
By contrast, the persistence of the transport at $7$~K is rather surprising, especially because 
carrier localization is expected to be enhanced  at lower temperatures.

The effect of the magnetic field up to $9$~T on the IX transport appears to be very weak. 
From the color-encoded spectra in Figs.~\ref{fig:transport7K} and \ref{fig:transport14K} one can see that the FWHM of the emission pattern does not change. 
This behaviour differs significantly from that of IXs in GaAs/(Al,Ga)As coupled QWs, where   magnetic field of the similar strengths has been shown to strongly inhibits IX transport.   \cite{Kuznetsova2017,Dorow2017}. 
This is due to the fact  that  IXs hosted by GaAs/(Al,Ga)As coupled QWs are much lighter particles, with much wider radii.
Therefore, the ratio $\chi$ between cyclotron energy $\hbar \omega_c$ and exciton binding energy $Ry$, $\chi=\hbar \omega_c /2 Ry $  increases much faster with magnetic field. 
%
%
When magnetic energy dominates over  the electron-hole Coulomb interaction, $\chi \gg 1$, the system is described in terms of so-called magnetoexciton \cite{Lozovik1997,Butov2001,Lozovik2002}.
The magnetic field leads in this case to the strong reduction of the in-plane Bohr radius $a_B$ and enhancement of the in-plane exciton mass $M$\cite{Edelstein,Lozovik2002,Butov2001,Stepnicki2015,Arnardottir,Wilkes2017,Wilkes2016,Kuznetsova2017,Dorow2017}. 
Therefore, the reduction of the diffusion coefficient $D=l\sqrt{2k_B T/M}$  and the mobility $\mu_X=D/k_BT $ is observed \cite{Kuznetsova2017,Dorow2017}.
Here   $l$  is the exciton scattering length and $k_B$ is the Boltzmann constant.
In GaN/(Al,Ga)N structures studied in this work and up to $9$~T we seem to deal with the opposite limit: 
$\chi \ll 1$. 
Numerical modelling of the IX transport within drift-diffusion model (see Appendix)~\cite{Fedichkin2016,Chiaruttini2019} allows us to estimate that the enhancement of the exciton mass $M(B)/M$ does not exceed a few percents at $B<9$~T, in consistance with estimations that can be done in the $\chi<1$ regime \cite{Arseev1998,Wilkes2016} : 
\begin{equation}
M(B)/M=(1-42\mu (a_B/l_B)^4/ 16^2M )^{-1}.
\label{eq:mass}
\end{equation}
Here $l_B=\sqrt{\hbar/eB}$ is the magnetic length,  $e$ is the electron charge, $\hbar$ is the reduced Planck constant,  $\mu$ is the exciton in-plane reduced mass.
Thus, in the following we assume that the shift of the IX emission energy measured at the center of the exciton spot in the presence of the magnetic field is entirely local, that is not affected by the in-plane transport, in contrast to the case of IXs in GaAs/(Al,Ga)As QWs.

\subsection{Magnetic field-induced exciton energy shift}
\label{sec:diamag}
\begin{figure}
\includegraphics [width=1\columnwidth] {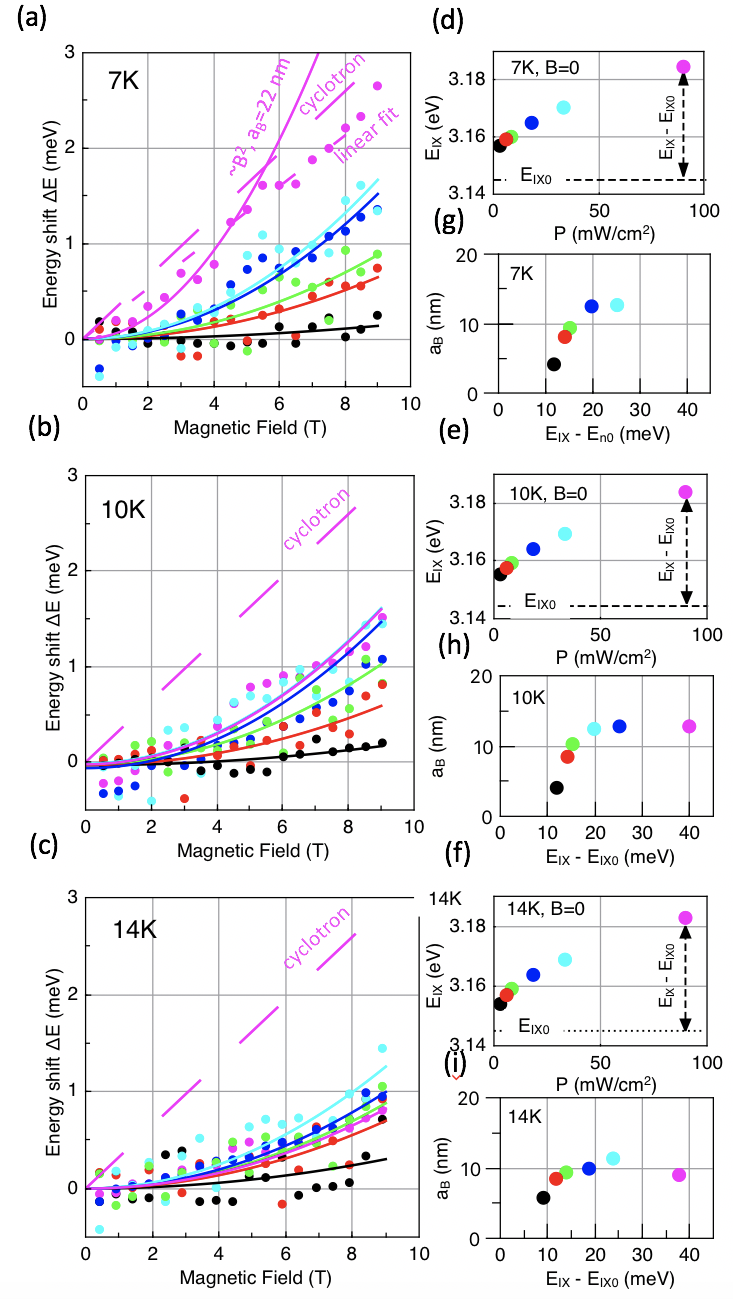}
  \caption{ (a-c) IX emission energy shift with respect to the $B=0$ emission energy 
  as a function of the magnetic field at $T=7$~K (a), $T=10$~K (b), $T=14$~K (c)  for various excitation power densities from $P=2.8$~mW/cm$^2$ (black symbols) to $P=90$~mW/cm$^2$ (magenta symbols). 
  Solid lines are parabolic fits to the diamagnetic shift model. The model does not fit the highest power data in (a), and the linear fit is shown by the long-dash line. Short-dash line in (a) shows the cyclotron shift expected for unbound electron-hole pairs.
 %
(d-f) show the emission energy at $B=0$ for each power density and three different temperatures. 
The horizontal dashed line shows zero-density limit $E_\mathrm{IX0}$ of the exciton emission energy. The vertical dashed arrow illustrates the density-induced blue shift of the emission line at $P=90$~mW/cm$^2$. 
 (g-i) Values of Bohr radius extracted from the quadratic fit (a-c)  as a function of the zero-field density-induced energy shift. 
Excitation power density is shown with the same colour code in (a-c), (d-f) and (g-i).  }
  \label{fig:diamag}
\end{figure} 
\begin{figure}
\includegraphics [width=0.9\columnwidth] {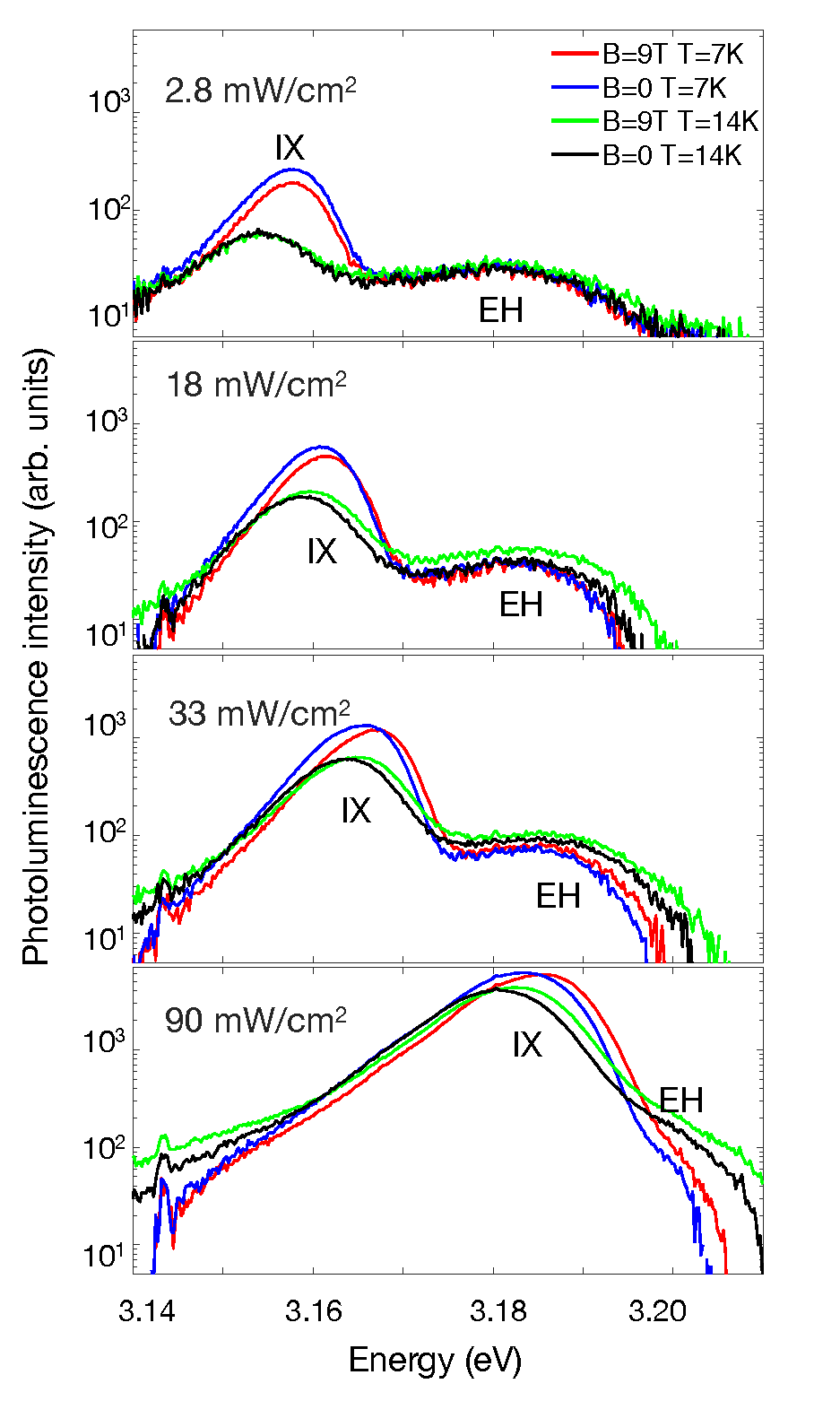}
  \caption{ PL spectra collected from $25$~$\mu$m-diameter area in the center of the excitation spot at different power densities. Two temperatures  ($T=7$~K, $T=14$~K) and two magnetic field values ($B=0$  and $B=9$~T) are shown for each power.  We identify the two emission lines as IXs and unbound electron-hole pairs (EH). }
  \label{fig:spectra}
\end{figure} 
The shift of the IX emission energy peak with respect to $B=0$ (averaged over $\approx 25$~$\mu$m) measured as a function of the magnetic field at different excitation power densities is shown by symbols in Fig. \ref{fig:diamag} for $T=7$~K (a), $T=10$~K (b), and $T=14$~K (c).
Panels (d-f) indicate for each temperature the correspondence between the emission energy $E_\mathrm{IX}$ measured at $B=0$ and the power density.
The colour code is the same as in (a-c).
The energies used to extract these shifts are obtained by the spectrum fitting procedure described in Appendix~\ref{par:lineshapeModelling}.
One can see that at maximum field $B=9$~T the energy shift varies  in the range from $0.2$ to $3$~meV. 
The largest shifts are achieved at highest powers and lowest temperatures.
Except for the highest power at $7$~K, the IX energy shifts  quadratically with the magnetic field. 
This can be interpreted as a diamagnetic shift $\Delta E=\gamma_D B^2$, where $\gamma_D=e^2 a_B^2/ (8\mu)$ is the so-called diamagnetic coefficient \cite{Bugajski}.
The diamagnetic coefficient appears to increase with  the excitation power, suggesting that IX Bohr radius increases as well.
This is a signature of  the progressive screening of the electron-hole Coulomb interaction within each IX.
The reduced exciton mass in GaN being well known  ($\mu=0.18 m_0$), fitting of the data to the quadratic dependence (solid lines in Fig.~\ref{fig:diamag}) allows us to extract the values of Bohr radii for all excitation power densities and temperatures.

The ensemble of the resulting values of $a_B$ are presented  in Fig. \ref{fig:diamag} (g-i). 
One can see that the Bohr radius can be as small as $5-6$~nm.
%
Note, that relatively small values of Bohr radius (compared to the prediction of the self-consistent  variational  calculation  of $a_B$ \cite{Bigenwald1999, Bigenwald2001}) are consistent with large exciton binding energy that we deduce from the PL spectra, where both exciton and free carrier emission are observed (see Fig.~\ref{fig:spectra}).  This will be further discussed in Sec.~\ref{sec:plplasma}. 

For each value of the sample temperature, the values of $a_B$ are shown as a function of the emission energy shift $E_{IX}-E_{IX0}$ at $B=0$.
This shift is related to the IX density in the QW and contains two contributions with different signs. 
The first one is the blue shift due to the interaction energy, and the second one is the red shift due to exciton binding energy (see Fig.~\ref{fig:fig1}).  
Their relative contribution will be discussed in Sec.~\ref{subs:density}.

Let us now  consider the set of measurements at $T=7$~K in Fig.~\ref{fig:diamag}~(a) that corresponds to the highest excitation power $P=90$~mW/cm$^2$ (magenta circles). 
In contrast with other measurements the energy shift does not grow quadratically with magnetic field. 
For comparison solid magenta line  in Fig.~\ref{fig:diamag}~(a) corresponds to $a_B=22$~nm, much larger than $a_B\approx12$~nm at higher temperatures and maximum powers. 
Instead, the data are much better described by a linear  function $\Delta E=\gamma_L B$, with the slope $\gamma_L=0.27$~meV/T .
Remarkably, the slope obtained from the linear fit is very close to  $\gamma_c=e\hbar/(2\mu)=0.32$~meV/T, that one could expect for the cyclotron energy of the electron-hole pair.
This suggests that, indeed, the density of the electron-hole pairs in the QW plane is high enough to screen the  Coulomb interaction within IX, so that the Mott transition could be reached. 

The reason why this transition could only be observed at $7$~K needs to be elucidated. 
In particular, it is important to find out wether this is due to the higher density reached in this case ({\it e.g.}  due to different nonradiative losses), or due to the screening of the exciton binding, that would be more efficient at low temperature.
Below we try to answer this question and  to determine  $n_\mathrm{Mott}$, the  critical density  at the Mott transition.
We recall  that the analysis made in this work is based on the assumption that the magnetic field up to $9$~T (that induces a cyclotron shift of the electron-hole pair emission $\hbar \omega_c/2=3$~meV ) can be considered as a small perturbation with respect to the exciton binding energy, {\it i.e.} $Ry \gg \hbar \omega_c/2$.
Therefore the formation of magnetoexciton does not take place.
This differs from the GaAs-based structures where the transition from Coulomb-bound exciton to magnetoexciton has been reported to occur at lower magnetic fields $\approx 2$~T \cite{Butov1999}.

\subsection{Photoluminescence of excitons {\it {vs}} free arriers}
\label{sec:plplasma}
Let us consider in detail the PL spectra collected from the $25$~$\mu$m-diameter area in the center of the excitation spot at different power densities and temperatures (Fig. \ref{fig:spectra}).
%
%
For the lowest excitation density (solid lines) we identify two emission lines, separated by $\approx 25$~meV,  the lowest energy line is the one that we have identified as the IX emission. 
It has  higher intensity and it is shifted towards higher energy by the magnetic field (as shown in Fig. \ref{fig:diamag}), while the second line is so wide, that it is difficult to quantify the energy shift induced by the magnetic field. 
When the excitation power increases, the IX emission strongly shifts towards higher energies up to merging with the higher energy emission line, which shifts much less.
Note also that the ratio between the intensities of the IX and the high energy line is $\approx5$ times higher at $T=7$~K than at $T=14$~K. 

We tentatively interpret the higher energy line as an emission of the unbound electron-hole (EH) pairs, that coexists with IXs even at the lowest excitation densities explored in this work. 
This contribution is not negligible as compared to the lower energy IX, because the corresponding density of states is huge.
EH emission can be spectrally separated from IX, because the linewidth of IX emission is smaller than IX binding energy.
Similar PL behavior has already been observed in both GaAs and InGaAs QWs \cite{Deveaud2005,Kappei2005,Kirsanske2016}. 
In GaN-based heterostructures the efficient formation of excitons even within a dense electron-hole plasma has already been observed by time-resolved terahertz spectroscopy \cite{Hangleiter2015}.
Both energy and intensity of the EH emission with respect to the well-identified IX emission are consistent with this interpretation: IX line merges with the EH plasma at high density, and the ratio between IX and EH intensity decreases when the temperature increases. 
The power-induced shift of the IX line is larger than that of the EH line. This is because EH transition energy is affected only by the screening of the built-in electrostatic potential, while IXs experience both this screening and the reduction of the binding energy.
Note also, that the EH emission line is much broader ($30$~meV at the lowest power, compared to $10$~meV for IXs) and weaker ($\approx 5$ times for the lowest power density) than the IX line.
That is why it was not possible to quantify the magnetic field-induced shift of the EH line, that is expected to be smaller than $3$~meV at highest accessible magnetic field.

The simultaneous observation of EH and IX emission provides us with some important information.
Indeed, the energy splitting between IX and EH lines characterises the exciton binding energy $Ry$, which, due to many-body effects, depends on the density of the particles in the QW \cite{ZimmermannBook}. 
From the measurements at the lowest power we estimate $Ry_0=25$~meV, corresponding to the zero-density limit. 
Exciton binding energy is related to its Bohr radius $Ry=e^2/4 \pi \varepsilon a_B$, where
$\varepsilon$ is dielectric permittivity constant in  the hosting semiconductor (here GaN). 
This yields $a_{B0}=6$~nm in the zero-density limit.
As already mentioned in the Sec.~\ref{sec:diamag}, these values are not properly described by the self-consistent  variational  calculation of the excitonic properties  \cite{Bigenwald1999, Bigenwald2000, Bigenwald2001}.
The relevance of this model relies on the choice of the variational function, and so far has not been verified experimentally.
Moreover, the model predicts that IX binding energy and Bohr radius remain constant up to 
approximately $10^{11}$~cm$^{-2}$ pair density (corresponding to emission energy shift $>10 $~meV), which also contradicts our experiments.
Thus, in what follows we rely on the experiments and use the value of $Ry_0=25$~meV and its relation to $a_B$ in order  to estimate the exciton density as a function of power and temperature and determine the Mott density. 
\section{Discussion}
\label{sec:discussion}
\subsection{Determination of the exciton densities}
\label{subs:density}
Estimation of the exciton densities and the critical densities for exciton dissociation is an extremely  delicate task, even in the case of traditional excitons with zero dipole moment. 
In particular, quantifying exchange and correlations within excitonic ensembles is a complex many-body problem. 
For IXs, density-dependent screening of the built-in electric field is another complicating factor, because it might be difficult to separate it from  exciton-exciton interaction.
We describe below the procedure that we adopted here to estimate power and temperature dependent exciton densities, Bohr radii and binding energies within available models. 
 
\begin{figure}
\includegraphics [width=0.9\columnwidth] {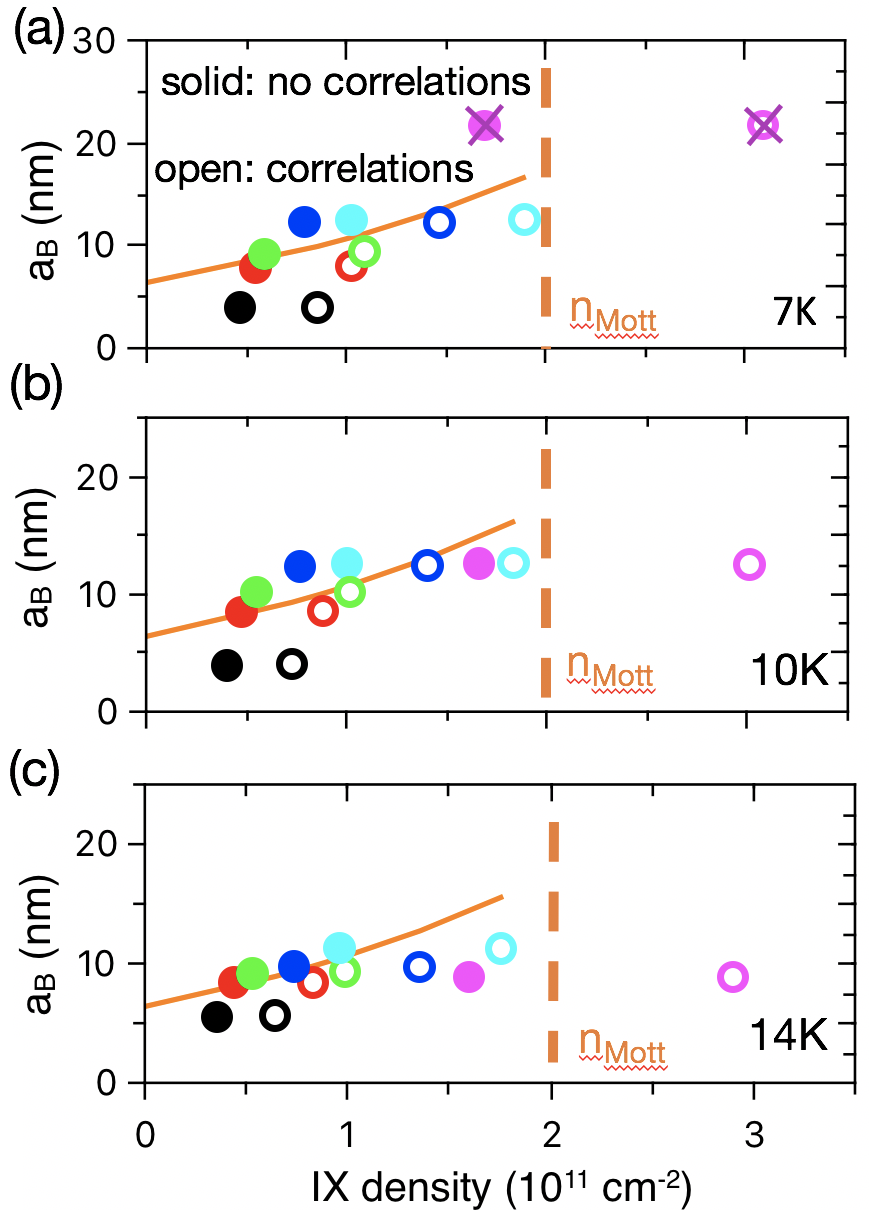}
  \caption{
Exciton Bohr radius (solid lines) as a function of  the IX density calculated within exponential model  assuming $n_\mathrm{Mott}=2\times10^{11}$~cm$^{-2}$ (dashed lines) and $Ry_0=25$~meV. These assumptions are identical for the three different temperatures (a-c). 
Symbols show the Bohr  radii extracted from the diamagnetic shift  at different powers (same as in Fig. \ref{fig:diamag}, same color code) as a function of the carrier density calculated  from the corresponding emission energy at $B=0$. Open (filled) symbols  show the calculation within Model C (Model NC).  At $7$~K crosses on top of the circles indicate the density points where magnetic field-induced energy shift is linear, so that the interpretation in terms of the diamagnetic shift is not satisfactory (the emission is expected to be dominated  by the electron-hole plasma). 
    }
  \label{fig:summary}
\end{figure} 

The   analysis is based on three experimentally determined (or deduced from them) parameters:  (i) the value of the zero-density binding energy $Ry_0=25$~meV estimated from the spectral splitting between IX and EH emission lines at the lowest density; (ii) the corresponding zero-density Bohr radius $a_{B0}=6$~nm; (iii) zero-density IX emission energy $E_\mathrm{IX0}=3.145$~meV, measured $\approx 700$~$\mu$m away from the excitation spot (see {\it e.g.} Figs. \ref{fig:transport7K}-\ref{fig:transport14K}~(c). 
These values are identical for all temperatures, within the accessible precision.

We start from the calculation of the electron-hole pair density corresponding to a given  band-to-band transition energy 
\begin{equation}
E_\mathrm{{EH}}(n)=E_\mathrm{{EH0}}+E_\mathrm{{int}}(n)
\label{eq:EEH}
\end{equation}
 using self-consistent solution of the coupled Schr\"{o}﻿﻿dinger and Poisson equations (see Appendix~\ref{par:SP}). 
The result of such calculation is very well approximated by the plate capacitor-like model, where density-induced blue shift is given by 
 $E_\mathrm{{int}}=\phi_0n$, with  $\phi_0=1.5\times10^{-13}$~eV cm$^{-2}$. 
Here $E_\mathrm{EH0}$ is the zero-density band-to-band transition energy, that does not yet account for the exciton binding energy $E_\mathrm{{EH0}}=E_\mathrm{{IX0}}+Ry_0=3.17$~eV.

As already mentioned, this approach completely neglects the  correlation effects which, at sufficiently high density and low temperature,  result in strong enhancement of the  exciton density expected at a given PL energy shift \cite{Zimmermann,Laikhtman}. 
We try to include these correlation effects in our model and  calculate $E_\mathrm{int}$ using a more sophisticated approach \cite{Zimmermann}.
Thus, we have two ways to calculate the band-to-band transition energy $E_\mathrm{EH}$, either including (Model C), or not including (Model NC) exciton correlation effects.
The details of these calculations are reported in the Appendix (Secs.~\ref{par:model_C} and \ref{par:SP}).

Because in our system $E_\mathrm{int}$ can reach values of the same order as $Ry_0$, that is  $\approx 10-20$~meV, it is important to take into account the variations of binding energy when the IX density is increased:  
\begin{equation}
E_\mathrm{IX}(n)=E_\mathrm{EH}(n)-Ry(n).
\label{eq:EIX}
\end{equation}
The simplest exponential model is chosen to account for the density-induced reduction of the exciton binding energy due to many-body effects:
\begin{equation}
Ry(n)=Ry_0 \times \mathrm{exp}(-n/n_\mathrm{Mott}),
\label{eq:Ry}
\end{equation}
where $n_\mathrm{Mott}$ is a  free parameter \cite{Liu2016}.
The corresponding Bohr radius $a_B(n)$ is calculated from $Ry(n)$ as :
\begin{equation}
a_B(n)=a_{B0} \times Ry(n)/Ry_0.
\label{eq:aB}
\end{equation}

Eqs.~\ref{eq:EEH} - \ref{eq:aB} can be solved for any value of the fitting parameter $n_\mathrm{Mott}$.  
This allows us to establish the relation between IX Bohr radius and particle density, either including (Model C) or not (Model NC) the excitonic correlations,  for each value of the excitation power and temperature.
The value $n_\mathrm{Mott}=(2\pm0.5) \times10^{11}$~cm$^{-2}$ ensures the best fit between the values of $a_B$ calculated from Eq.~\ref{eq:aB} and those extracted from magnetic field-induced diamagnetic shift.
%
%
%

%
The results of this analysis are summarized in Fig.~\ref{fig:summary} for three different temperatures. 
The correspondence  between IX density and Bohr radius calculated within  Model C (open symbols) and Model NC (filled symbols)  is shown for all excitation powers. 
At $7$~K crosses on top of the circles indicate the highest density point ($P=90$~W/cm$^2$), where the energy shift induced by magnetic field is linear, so that the interpretation in terms of  diamagnetic shift is not satisfactory. 
Lines indicate the model assumption for $a_B(n)$  given by Eq.\ref{eq:aB} (solid line) and Mott density resulting from the best fit to the data (dashed line).

One can see that within the relevant density/temperature parameter range the difference in the density estimation between two models does not exceed a factor of two. 
At all powers except the highest one (magenta symbols), both models  describe the experimental data reasonably well. 
Namely, exciton density remains below $n_\mathrm{Mott}$, consistent with the experimentally observed diamagnetic behaviour. 
By contrast, at $P=90$~mW/cm$^2$ only Model C allows us to describe the Mott transition observed at $T=7$~K.
Nevertheless, despite the temperature dependence that it intrinsically includes (see Appendix), Model C fails to describe the absence of Mott transition at $T=10$ and $14$~K.
It seems to overestimate the role of the correlations at $T>10$~K, while simpler Model NC provides better agreement with the data.
Thus, none of  the models fits to the ensemble of the data. 
In the attempt to reconstruct the IX phase diagram shown in Fig.~\ref{fig:fig1}, we assume that the densities are given by Model C at  $T=7$~K and by Model NC at higher temperatures.

\subsection{IX phase diagram}
\label{sec:phase}
The  determination of the exciton densities and the Mott transition (only reached at $T=7$~K) allows us to 
draw the parameter space explored in this work on the theoretical IX phase diagram presented in Fig.~\ref{fig:fig1}~(a).
The critical temperature for the formation of the BEC-like state is given by  $k_B T_\mathrm{{BEC}}=2\pi \hbar^2 n/M$ (red area), and  for the formation of the correlated dipolar liquid $k_B T_\mathrm{{DL}}=(4\pi\varepsilon n/(ed)^2)^{2/3}$ (magenta area) \cite{Laikhtman,Butov2016,Combescot2017}.
Here $\varepsilon$ is the dielectric constant in the matter (GaN).
The exciton parameter space explored here (grey area) is expected to span over three different phases:  gas, 
dipolar liquid and electron-hole plasma.
%

The Mott transition from excitons to ionised electron-hole pairs at $n_\mathrm{Mott}\approx2\times10^{11}$~cm$^{-2}$ has only been observed at $7$~K, because only at this lowest explored temperature such high density could be reached. 
Indeed, the experimental configuration that we used relies on broad excitation, that limits accessible laser power density. 
Some  non-radiative losses could be sufficient to reduce the carrier density by a factor of the order of two, making the transition inaccessible.
Since we have no experimental determination of the temperature dependence of the Mott density, it is represented in Fig.~\ref{fig:fig1}~(a) as  a temperature-independent phase boundary. 
Note that in narrow GaN/(Al,Ga)N QWs hosting excitons such that their dipolar moment is negligibly small and  $n_\mathrm{Mott}$ is  three times higher than in our structure,  no temperature dependence of the Mott density have been found up to $150$~K \cite{Rossbach2014}.
The estimated value of $n_\mathrm{Mott}$ is three times lower than the one reported in similar structures but with narrow QWs, and thus non-dipolar excitons \cite{Rossbach2014}. 

The interpretation of the Mott transition observed at $7$~K is complex. 
Our analysis of the underlying densities as well as the theoretically expected formation of the liquid stated suggests the importance of the excitonic correlations at this temperature. 
Another indication on the onset of the correlations, is the differences in the  excitonic transport at $7$~K and  at higher temperatures. 
As pointed out in Sec. \ref{sec:transport} (Fig.~\ref{fig:transport7K}), at $7$~K the excitonic drift remains efficient  even at the lowest density used in this work. 
This behaviour is not correctly described by the simple transport model that does fit properly the transport at $14$~K. 
Note, that the enhancement of the exciton propagation at low temperature exclude the localization effects as a possible cause of the observed effects.

%

The precision of the procedure used in this work to explore the onset of the many-body effects has of course a number of limitations.
In particular, we  use an oversimplified relation between exciton Bohr radius and its binding energy and neglect non-radiative effects.
Moreover, our  considerations neglect possible magnetic field dependence of the exciton Bohr radius and eventual transition towards magnetoexiton states at highest magnetic fields \cite{Butov1999}.
Although electron-hole binding  energy seems to remain higher than the cyclotron energy these effects may affect the results. 
To confirm and better quantify the onset of many-body effects (dipole-dipole excitonic correlations and Mott transition)  future work should focus on (i) lowering down the temperature of the carriers in this system, (ii) comparing the systems hosting IXs with different dipole length.
Measurable effects can be expected already at $T=2$~K, and with a $2-3$~nm variation of the QW width.

\section{Conclusions}
\label{sec:conclusions}
In conclusion, using the PL spectroscopy in Faraday configuration we demonstrated the transition from the dipolar IX fluid to the EH plasma in wide polar GaN/(AlGa)N QWs at $7$~K.
The corresponding pair density $n_\mathrm{Mott}=(2\pm0.5)\times10^{11}$~cm$^{-2}$ could not be reached at higher temperatures, presumably due to non-radiative losses.
We have shown that below Mott transition, IX fluid and  unbound EH pairs coexist and can be spectrally separated. 
The increase of carrier density in the system is accompanied by the screening of both the built-in electric field and the IX binding energy. 
In the explored parameter range, these two contributions are of the same order.
Our modelling suggests  the build-up of the exciton-exciton correlations and the possible 
formation of the dipolar liquid at lowest temperature $T=7$~K.
Finally, in contrast with GaAs-hosted IXs, the radial transport of the IXs GaN/(Al,Ga)N QWs appears to be preserved under magnetic field up to $B=9$~T. 

\renewcommand{\arraystretch}{1.5}
\begin{table}[h!]
\begin{center}
\begin{tabular}{lcc}
\toprule
\hline
 Parameter & Units &Value \\
 \hline
\midrule
QW width & (nm) & $7.8$ \\
Al composition in the (Al,Ga)N barriers& (\%) & $11$ \\
Al$_{0.11}$Ga$_{0.89}$N cap layer thickness & (nm) & 50 \\
Al$_{0.11}$Ga$_{0.89}$N  bottom layer thickness & (nm) & 100 \\
GaN buffer layer thickness  & (nm) & 800 \\
Dislocation density in GaN substrate& (cm$^{-2}$) & $\sim 2\times 10^7$ \\
IX emission energy $E_{IX0}$ at $n\approx 0$    & (eV)       & $ \sim 3.145$ \\
IX binding energy $Ry_{0}$ at $n\approx 0$    & (meV)       & $ \sim 25$ \\
IX radiative lifetime  $\tau_0$ at $n\approx 0$ & ($\mu$s) & $\sim 40$ \\
$m^z_{e}$ for electron in GaN    & ($m_0$)& $0.200$\\
$m^z_{h}$ for hole in GaN & ($m_0$)& $1.1$\\
$m^z_{e}$ for electron in Al$_{0.11}$Ga$_{0.89}$N    & ($m_0$)& $0.213$\\
$m^z_{h}$ for hole in Al$_{0.11}$Ga$_{0.89}$N & ($m_0$)& $1.367$\\
$m^\perp_e$ transverse electron mass in QW & ($m_0$)& $0.2$\\
$m^\perp_h$ transverse hole mass in QW & ($m_0$)& $1.6$\\
$\varepsilon$ Dielectric constant in GaN & ($\varepsilon_0$) & $8.9$\\
$\varepsilon_B$ Dielectric constant in (Al,Ga)N & ($\varepsilon_0$) & $8.926$\\
GaN bandgap energy at 7~K & (eV)& $3.507$\\

(Al,Ga)N bandgap energy at 7~K& (eV)& $3.709$\\
Band offset in GaN and (Al,Ga)N & (\%) & $80$\\
\bottomrule
\end{tabular}
\end{center}
\caption{Sample and material parameters used in various calculations \cite{Vurgaftman2001,Vurgaftman2003,Kabi,Rickert2002}.}
\label{table:sampleData}
\end{table}

\renewcommand{\arraystretch}{1.5}
\begin{table}[h!]
\begin{center}
\begin{tabular}{lcc}
\toprule
\hline
Parameter   & Units &Value \\
\midrule
\hline
Built-in electric field $F$ & (kV/cm)  & 980 \\
Exciton radiative lifetime factor $\gamma$ & (cm$^{-2}$) & $2.7 \times 10^{11}$ \\
IX-IX interaction constant $\phi_0$ & (eV$\cdot$cm$^{2}$) & $1.50\times 10^{-13} $\\ 
IX Bohr radius $a_{B0}$ at $n\approx 0$    & ($\mu$m)       & $ \sim 6$ \\
Band-to-band transition  $E_{EH0}$   & (eV)       & $ \sim 3.17$ \\
\bottomrule
\end{tabular}
\end{center}
\caption{Sample parameters from self-consistent solution of Schr\"{o}dinger and Poisson equations.}
\label{table:calcData}
\end{table}


\section{Acknowledgments} 
We are grateful to L. V. Butov for enlightening discussions.
This work was supported by  the 
French National Research Agency via OBELIX and IXTASE projects, as well as LABEX GANEX. 
\section{Appendix}
 \label{sec:appendix}
\setcounter{equation}{0}
\renewcommand{\theequation}{A\arabic{equation}}
\renewcommand{\thesubsection}{A\arabic{subsection}}
\setcounter{figure}{0} \renewcommand{\thefigure}{A.\arabic{figure}}
\setcounter{table}{0} \renewcommand{\thetable}{A.\arabic{table}}

\subsection{Sample structure}
\label{par:sampleStructure}
The sample is grown by molecular beam epitaxy (MBE) on a \textit{c}-plane oriented n-type GaN LUMILOG substrate with $\sim 2\times 10^7 \mathrm{cm^{-2}}$ threading dislocation density. The substrate is covered  by a $0.8$~$\mu$m-thick GaN buffer layer under the active zone. The active zone consists of a $7.8$~nm-wide GaN QW sandwiched between $50$~nm (top) and $100$~nm-wide (bottom) Al$_{0.11}$Ga$_{0.89}$N barriers.
%
%
The main characteristics of the sample, material properties used in calculations, and the results of various calculations are summarized in Tables~\ref{table:sampleData}-\ref{table:calcData}.
%


\subsection{Spatially-resolved PL setup under magnetic field}
\label{par:PLsetup}

The sample is placed in the chamber of a variable temperature insert (VTI) of the magneto-optical cryostat (\textsc{Cryogenic Limited}). 
The magnetic field up to $9$~T is created by superconducting coils in Faraday geometry.
 
\begin{figure}[h]
    \includegraphics[width=1\columnwidth]
{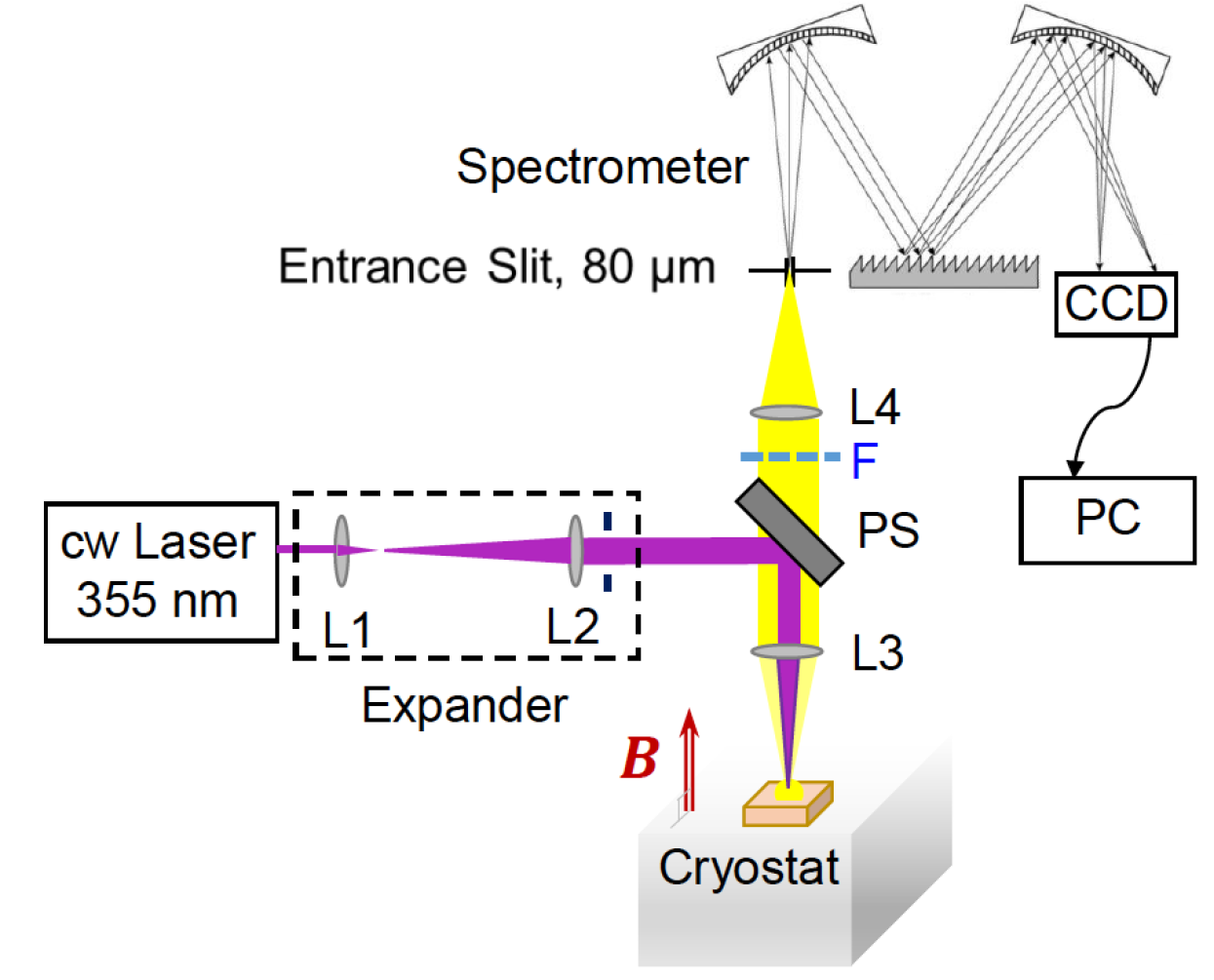}
    \caption{Experimental set-up.}
    \label{fig:principleBfield}
\end{figure}

Optical set-up is shown in the Fig.~\ref{fig:principleBfield}. The laser source is a continuous emission at 355~nm (Coherent OBIS, nominal power 20~mW). The beam is first shaped by an expander (L1, L2) to increase the numerical aperture incident on the objective lens (LO) which focuses the excitation on the sample surface. The diameter of the laser spot thus focused can be reduced upto about 20~\textmu m. We deliberately deregulated the beam expander in order to increase the size of the incident spot on the sample up to  300~$\mu$m.
The luminescence is collected by the same objective lens (L3). A bandpass filter (F, Semrock 400/16~nm) placed behind a power-splitter (PS) eliminates the reflection of the laser beam. The sample is imaged on the spectrometer (Horiba Jobin Yvon iHR550) entrance slit (80~$\mu$m) . The signal is then sent to a CCD~ camera: each pixel corresponds to 27~\textmu m on the sample.

\subsection{Modelling of the IX transport}
\label{par:IX_transport}
The model used here is identical to the one that we used in Refs.~\onlinecite{Fedichkin2016,Chiaruttini2019}.
It is based on the equation for the in-plane transport of indirect excitons.
 In its most general form reads:\cite{Ivanov2002,Rapaport2006}
\begin{equation}
\frac{\partial n}{\partial t}=
- \nabla \cdot \mathbf{J} + G - R n,
\label{eq:drift-diffu}
\end{equation}
where $n$ is the exciton density,
\begin{equation}
G= \frac{N_p}{\left( \pi a_0^2 + \pi w^2/2 \right )} \times \textrm{erfc}\left( \frac{r-a_0}{w}  \right)
\label{eq:G}
\end{equation}
 is the exciton density generation rate,  $N_p$ is the number of excitons generated per second, $a_0$  and $w$ are the laser spot geometric factors, $R$ is the recombination rate, $\mathbf{J}$ is the IX current density.
In order to describe the spatial profile of the spot we have chosen the parameters in Eq.~\ref{eq:G} as $w=100$~$\mu$m, $a_0=200$~$\mu$m. 
%
%
The exciton current $\mathbf{J}$ can be split into drift and diffusion components:
$\mathbf{J}= \mathbf{J}_\mathrm{drift} + \mathbf{J}_\mathrm{diff}$.

The diffusion current $\mathbf{J}_\mathrm{diff}$ is given by:
\begin{equation}
\mathbf{J}_\mathrm{diff}=
-D \nabla n
\label {eq:diffcurr}
\end{equation}
where $D$ is the exciton diffusion coefficient $D=l\sqrt{2k_B T/M}$.
It depends on the exciton scattering length  $l$ \cite{Rapaport2006,Fedichkin2016}, but also on the exciton in-plane mass $M=m^\perp_e+m^\perp_h$ that can be affected by the magnetic field: $M(B)/M=1+\delta M$. 
%
%
%
The drift term $\mathbf{J}_\mathrm{drift}$ in Eq.~\ref{eq:drift-diffu} can be
rewritten as
\begin{equation}
\mathbf{J}_\mathrm{drift}=
-\mu_X n \nabla (\phi_0 n)
\label{eq:Jdrift}
\end{equation}
where the exciton mobility $\mu_X$ is connected to the diffusion coefficient $D$
via the  Einstein relation: $\mu_X=D/k_{B}T$.
In Eq.~\ref{eq:Jdrift}, the drift is governed by the exciton-exciton interaction energy $\phi_0 n$ only.
The recombination rate $R$ is assumed to be dominated by radiative mechanisms $R= 1/\tau_{rad}$, where 
density dependent radiative recombination time is given by $\tau_{rad}=\tau_0 \exp ( - n/\gamma)$  and $\gamma$ is obtained from the the solution of the coupled Schr\"odinger and Poisson equations \cite{Lefebvre2004,Fedichkin2016,Chiaruttini2019} and $\tau_0=40$~ms (see Table \ref{table:calcData} ).   
%

We look for the steady-state solutions $n(r)$  of   Eq. (\ref{eq:drift-diffu})  at $T=14$~K in order to evaluate the eventual effect of the IX mass enhancement on the in-plane transport.  
The emission profiles measured  experimentally are  compared with the calculated spatial profiles of (i) the exciton emission energy shift $\Phi_0 n$  and (ii) the exciton PL intensity $I=n R$ at two different power densities, the lowest and the highest ones. 
The corresponding values of   $N_P$ in Eq.~\ref{eq:G} are $N_P=1.2\times10^{12}$~s$^{-1}$ for $P=2.8$~mW/cm$^2$ and $N_P=40\times10^{12}$~s$^{-1}$ for $P=90$~mW/cm$^2$. 
%
%

%
%
\begin{figure} [h]
\includegraphics [width=1\columnwidth]{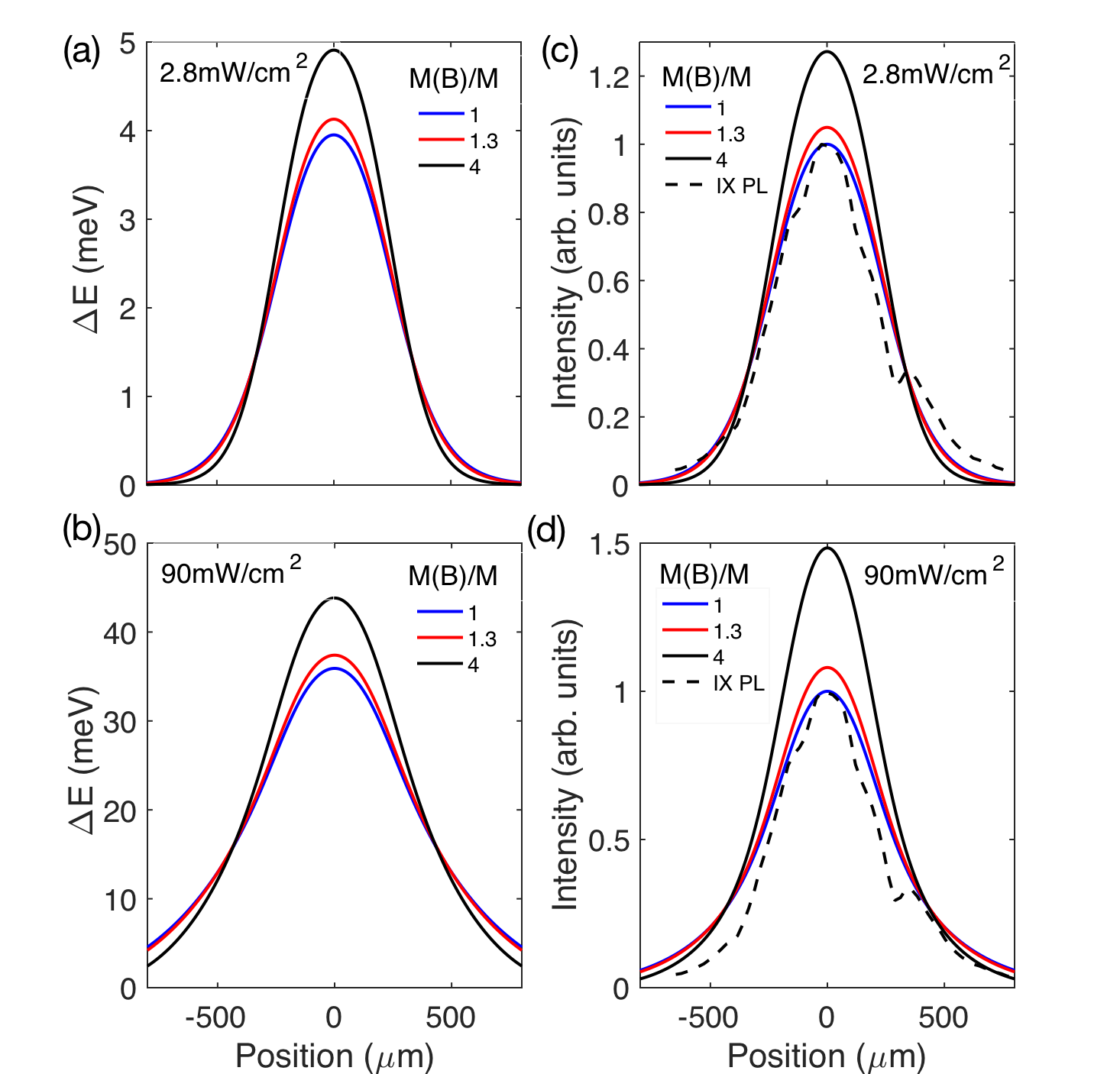}
  \caption{ Calculated spatial distribution of the emission energy shift (a-b) and  the intensity (c-d) for three different values $M(B)/M$ and two power densities. Dashed lines in (c-d) show measured IX intensity at $T=14$~K (there is no difference in the $B=0$ and $B=9$~T profiles).
  }
\label{fig:profileB}
\end{figure} 
\begin{figure} [h]
\includegraphics [width=1\columnwidth]{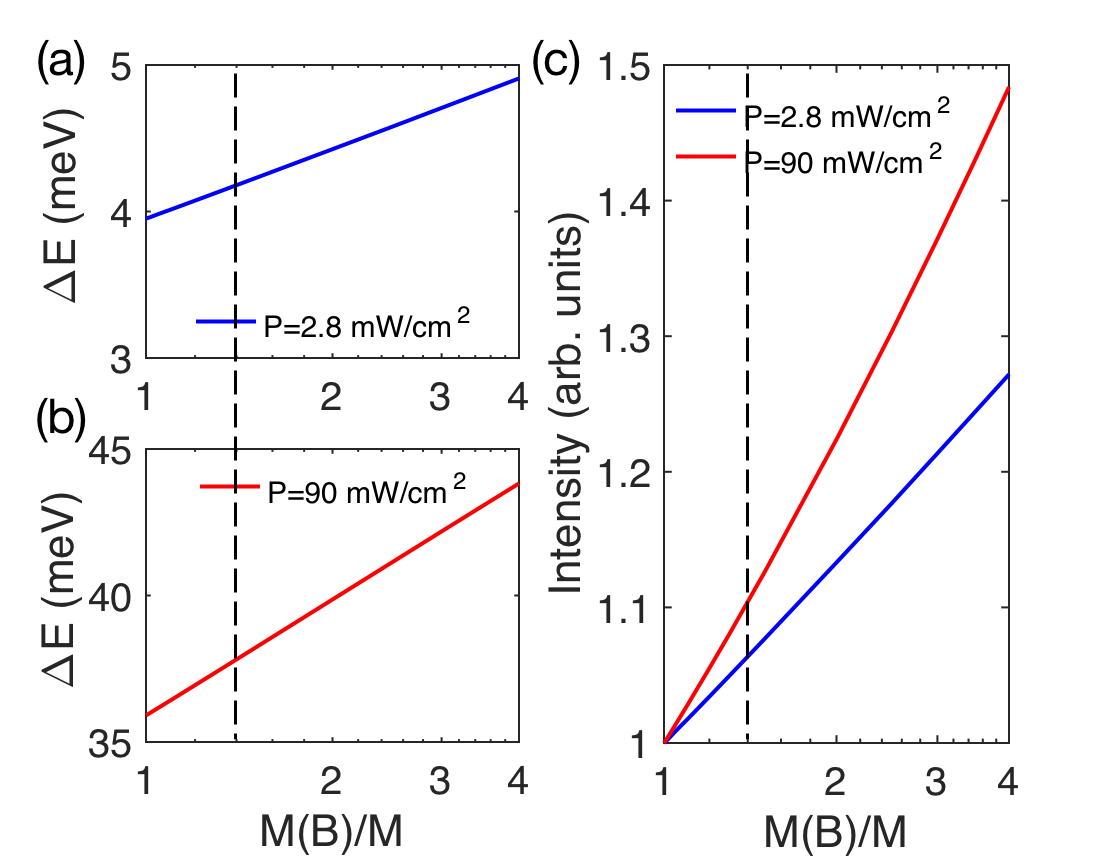}
 \caption{Energy shift (a-b) and intensity increase (c) in the center of the excitation spot  calculated as a function of $M(B)/M$ for  two different values  of power density. Dashed lines point the value  $M(B)/M=1.3$, the estimated limit for our system.
  }
\label{fig:shiftB}
\end{figure} 

Fig.~\ref{fig:profileB}  shows how the emission energy shift and  the intensity profiles calculated for three different values $M(B)/M$ and two power densities.
One can see that while the modification of the emission profile would be hardly detectable experimentally even for  $M(B)/M=4$, the energy  shift in the spot centre is sensible to the mass enhancement, and is accompanied by the  significant enhancement of the emission intensity.
This effect is strongest at high power.

Fig.~\ref{fig:shiftB} shows the calculated energy  shift and emission intensity in the spot centre as a function of $M(B)/M$. 
Since experimentally we do not observe any detectable increase of the emission intensity at $B=9$~T 
with respect to zero field emission, we estimate that the corresponding mass enhancement does not exceed $M(B)/M<1.3$ (dashed lines in Fig.~\ref{fig:shiftB}.
Assuming that  $M(B)/M$ is described by Eq.~\ref{eq:mass}, we obtain that $a_B<16$~nm, which is consistent with our results.
\subsection{Modelling of the PL lineshape}
\label{par:lineshapeModelling}

The spectral profile of the excitonic luminescence have in semi-logarithmic representation
%
 a triangular shape whose low (high) energy tails can be modelled by an exponential $\exp(\pm \beta_{1(2)} E)$,
where $\beta_{1(2)}$ characterises the slope of the low (high) energy tail.

\begin{figure} [h]
\includegraphics [width=1\columnwidth]
{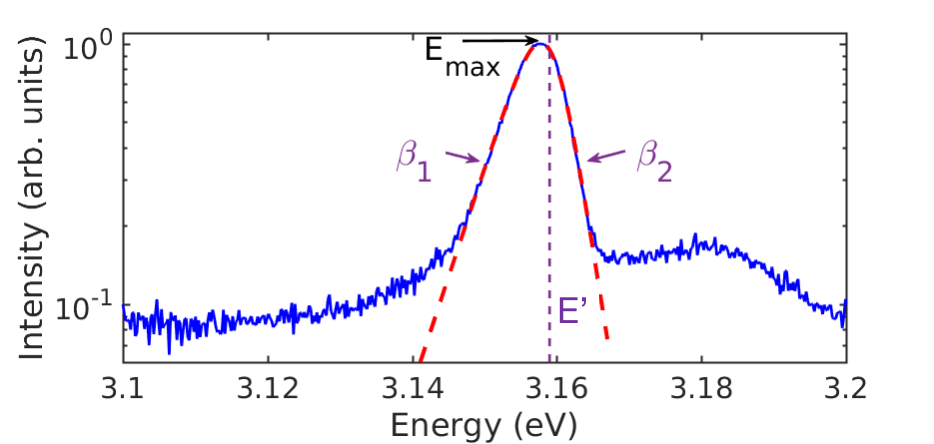}
\vspace{-0.4cm}
  \caption{A typoical IX photoluminescence spectrum (blue) and the result of the fitting procedure (red dashed). 
  }
  \label{fig:fitProcedure2020_log1}
\end{figure} 

We use the following generic function to describe the the IX spectrum:
\begin{equation}
 I=f_\theta(E) = A \frac{\exp(\beta_1(E-E'))}{1+\exp(\beta_2(E-E'))}
\end{equation}
where $\theta=\{A,\beta_1,E',\beta_2\}$ is the set of parameters to be optimised. 
Our algorithm  employs a least squares regression from a suitably predetermined initial condition.
Note that the parameter $E'$ does not correspond to the spectral maximum, $\mathrm{E}_\mathrm{max}$: the latter is thus obtained by a simple numerical search of the maximum value on a list of energy values applied to the model function $x \to f_{\theta_0}(x)$ where $\theta_0$ is the optimum in the sense of least squares.

Figure~\ref{fig:fitProcedure2020_log1} illustrates the relevance of the chosen fitting function and the quality of the line shape modelling.

\subsection{Estimation of density-dependent exciton energy shift.}
\label{sec:Adensity}
We use two  models to establish the relation between exciton density and the emission energy shift.
The first model (Model NC, Sec. \ref{par:SP} ) is based on the solution of  the coupled Schr\"odinger and Poisson  equations \cite{Lefebvre2004}. 
It accounts for the screening of the built-in electric field by the photocreated carriers, but neglects the correlations that can settle between them at low temperature and high densities.
The second model (Model C, Sec. \ref{par:model_C}) accounts for both screening and correlations.
 Both models are described below. 

\subsubsection{Self-consistent solution of SP equations (NC model)}
\label{par:SP}

Electron and hole wavefunctions are calculated by solving the 1D Schr\"{o}dinger equation for the envelope function, 
$\Psi(z)$:
\begin{align}
 -\frac{\hbar^2}{2}\frac{\mathrm{d}}{\mathrm{d}z}\left( \frac{1}{m^z(z)} \frac{\mathrm{d}}{\mathrm{d}z} \Psi(z) \right)+V(z) \Psi(z)=E\Psi(z).
\end{align}
Here $V(z)$ ($m^z(z)$) stands for the corresponding band potential (effective mass).
We implement a second order finite difference scheme of uniform spatial step, $\Delta z$, that leads to the following secular equation
\begin{align}
 \left(H_{ij} +V_i\delta_{ij}\right) \Psi_j=E_j\Psi_j.
\end{align}
Here $H\equiv(H_{ij})$ is a $N\times N$ tridiagonal matrix given by~:
\begin{align}
H=\left( \begin{array}{ccccc} D_1&U_1&&\dots&0\\
L_2&D_2&U_2&\dots&0\\ \vdots&\ddots &\ddots &\ddots&\vdots \\ & & & &U_{N-1}\\
0&\dots&0&L_N & D_N \end{array}\right).
\end{align}
Here $N$ is the number of nodes in the grid, while upper ($U_i$), lower ($L_i$) and diagonal ($D_i$) coefficients are given, for $1<i<N$, by
\begin{align}
 U_{i}=&\frac{\hbar^2}{4\Delta z^2}\left(\frac{1}{m_{i+1}}+
 \frac{1}{m_{i}}\right)\\
 L_{i}=&\frac{\hbar^2}{4\Delta z^2}\left(\frac{1}{m_{i}}+
 \frac{1}{m_{i-1}}\right)\\
 D_i=&-\frac{\hbar^2}{4\Delta z^2}\left(\frac{1}{m_{i+1}}
 +\frac{2}{m_{i}}+\frac{1}{m_{i-1}} \right)
\end{align}
Using $\Delta z \sim 0.1$~nm,  the eigenvalues are computed with $\sim10^{-6}$ relative precision.
By applying the above formalism to both conduction and valence bands we obtain the corresponding ground state confinement energies.
Their difference gives us the energy of the fundamental optical transition $E_{EH0}$.

Note, that we implement the simplest Born-Von Karman cyclic boundary conditions, that imply that the band potential is identical at the two boundaries of the active layer \cite{Gil2014}. 
One must bear in mind, however, that in reality the situation is more complex, and the potential drops across the active layer. Fermi level is known to be pinned at the surface, where  a depletion region forms \cite{Segev2006}; a two-dimensional electron gas resulting from residual doping and surface states accumulates on the back side of the active structure (at the bottom (Al,Ga)N/GaN  interface).  
We checked that, for this sample structure, the Fermi level is well below the bottom of the QW~\cite{Chiaruttini2019}, which means that one should not expect the presence of the residual carriers in the active layer.

\begin{figure}[h!]
\centering
\includegraphics [width=0.75\columnwidth] {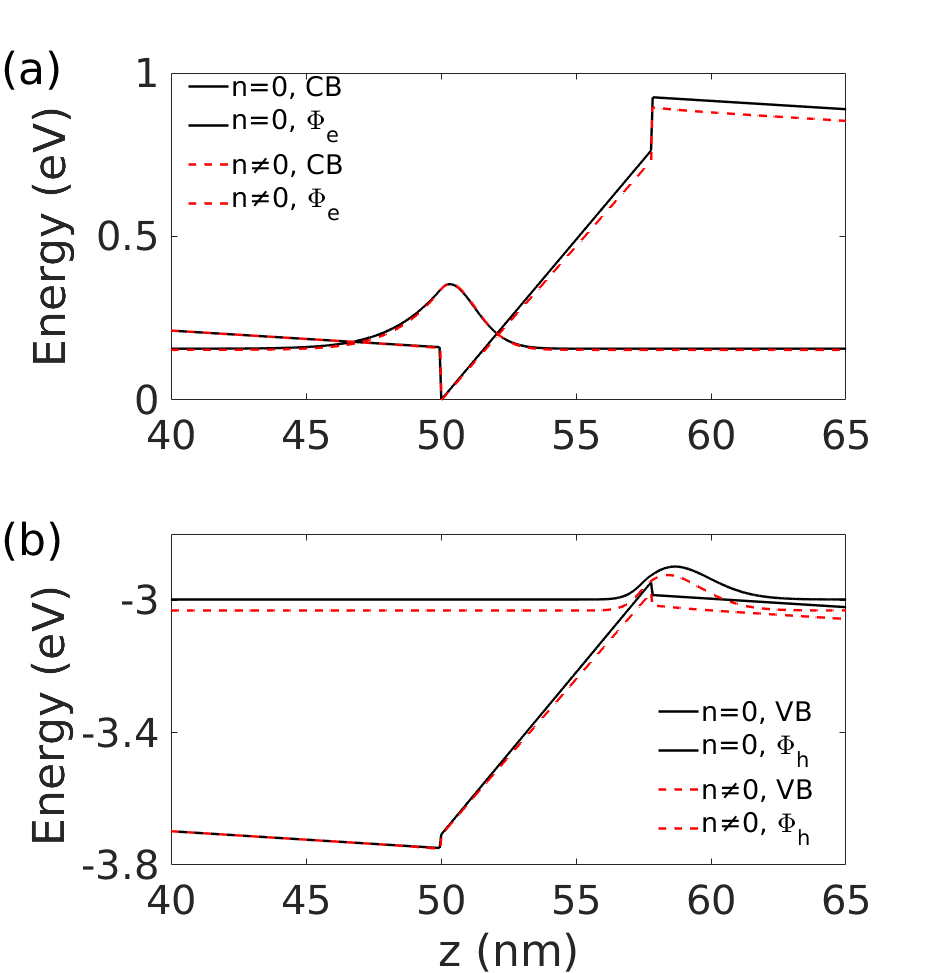}
\vspace{-0.3cm}
  \caption{Effect of the photoinjected carriers on the conduction (a) and valence (b) band diagram and wavefunction.
  The band profiles and wavefunctions are calculated for an empty QW (black lines) and for a pair density of $n=2 \times 10^{11}$~cm$^{-2}$ (red dashed lines).}
  \label{fig:densityInfluence}
\end{figure}
 \begin{figure}[]
 \centering
\includegraphics[width=1\columnwidth]{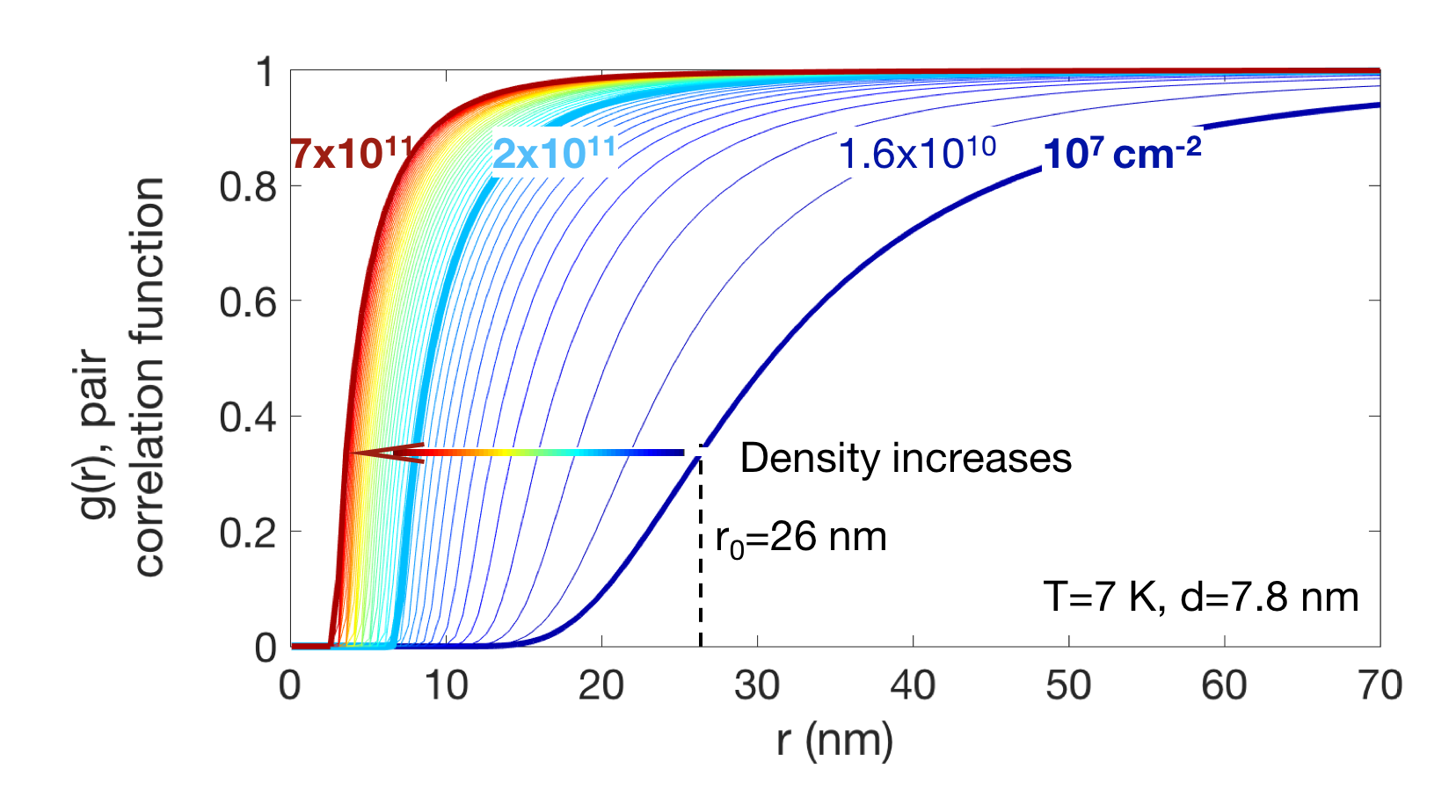}
  \caption
  { Pair correlation function calculated for various values of the particle density at $T=7$~K and dipole length $d=7.8$~nm. Thick lines indicate the calculations in the low density limit ($n=10^{7}$~cm$^{-2}$, dark blue), at Mott density estimated in this work ($n=2\times10^{11}$~cm$^{-2}$, cyan) and at  $n=7\times10^{11}$~cm$^{-2}$, corresponding to the saturation of $g(r)$. Dashed line poins $r_0$, the mean-field exclusion radius given by Eq.~\ref{eq:r0}.}
    \label{fig:paireCorrelationFonction}
\end{figure}
Photoinjected carriers affect the band potential that confines electrons and holes in the quantum well. For a given charge density profile $\rho(z)$ 
the resulting potential is given by the Poisson equation
\begin{equation}
 \frac{\mathrm{d}}{\mathrm{d}z} \left(\varepsilon(z) \frac{\mathrm{d}}{\mathrm{d}z} V(z)\right) = -\rho(z)
\end{equation}
where $\varepsilon(z)$ is the local dielectric permittivity.

Using boundary conditions at $z=0$  (nullity of both electric potential and field) yields:
\begin{equation}
 V(z)=-\int_0^z \mathrm{d}\xi \left[\frac{1}{\varepsilon(\xi)}\int_0^{\xi} \mathrm{d}\xi' \, \rho(\xi') \right].
\end{equation}
Thus, to account for the screening of the built-in potential by the photoinjected carriers  we solve Poisson and Schr\"{o}dinger
equations self-consistently.

An example of such calculation for the electron-hole pair  density $n=2\times10^{11}$~cm$^{-2}$ and for $n=0$ in our sample is shown in Fig.~\ref{fig:densityInfluence}. One can see that the band diagram is less tilted in the presence of the carriers, due to the electrostatic screening effect. This implies the modification of both optical transition energy and the overlap $\Omega=\left|\int \mathrm{d}z\,\Phi_e(z) \Phi_h(z)\right|^2$ between the electron and hole wavefunctions ($\Omega$  is proportional to the radiative emission rate).

The transition energy increases almost linearly with density (figure not shown), at least in the density limit of interest. The slope $\phi_0=1.50 \times 10^{-13}\textrm{ eV}\cdot \textrm{cm}^{2}$ is obtained from the linear fitting procedure.
 It differs only slightly from the value given by a simple "plate capacitor" model 
\begin{align}
 \phi_0^{PC}=\frac{e^2 d}{\varepsilon}\approx 1.59\, 10^{-13}\textrm{eV} \cdot \textrm{cm}^{2}
\end{align}
This difference is due to the localization of the electron and hole wavefunctions in the inner part of the QW, which tends to reduce the effective QW width. 
The interaction energy $E_{int}$ given by $E_{int}=\phi_0 n$ is shown in Fig. \ref{fig:EintComputed}
by the black line. 

\subsubsection{Taking into account excitonic correlations (Model C)}
\label{par:model_C}
A more accurate estimate of the excitonic density from the energy shift is based on taking into account the repulsive interactions between excitons.
This  means that one should go beyond the assumption of the  uniform distribution of charges in the plane of the quantum well,  on which the "plate capacitor" and the Schr\"{o}dinger-Poisson models are based.

The energy of an exciton in the field of its neighbors can be written as
\begin{equation}
 E_{\mathrm{int}}^\mathrm{corr}(n,T)=\int n_{\ell}(r,T,n)U(r) d \mathbf{r}
 \label{eq:nl} 
\end{equation}
 where $U(r)=\frac{2e^2}{4\pi\varepsilon}\left(\frac{1}{r}-\frac{1}{\sqrt{d^2+r^2}}\right)$ is the exciton-exciton interaction potential and $n_{\ell}(r,n,T)$ is the local density of excitons in equilibrium at temperature $T$ and for a global density $n$. 
 
\begin{figure}[h]
\includegraphics[width=1\columnwidth]{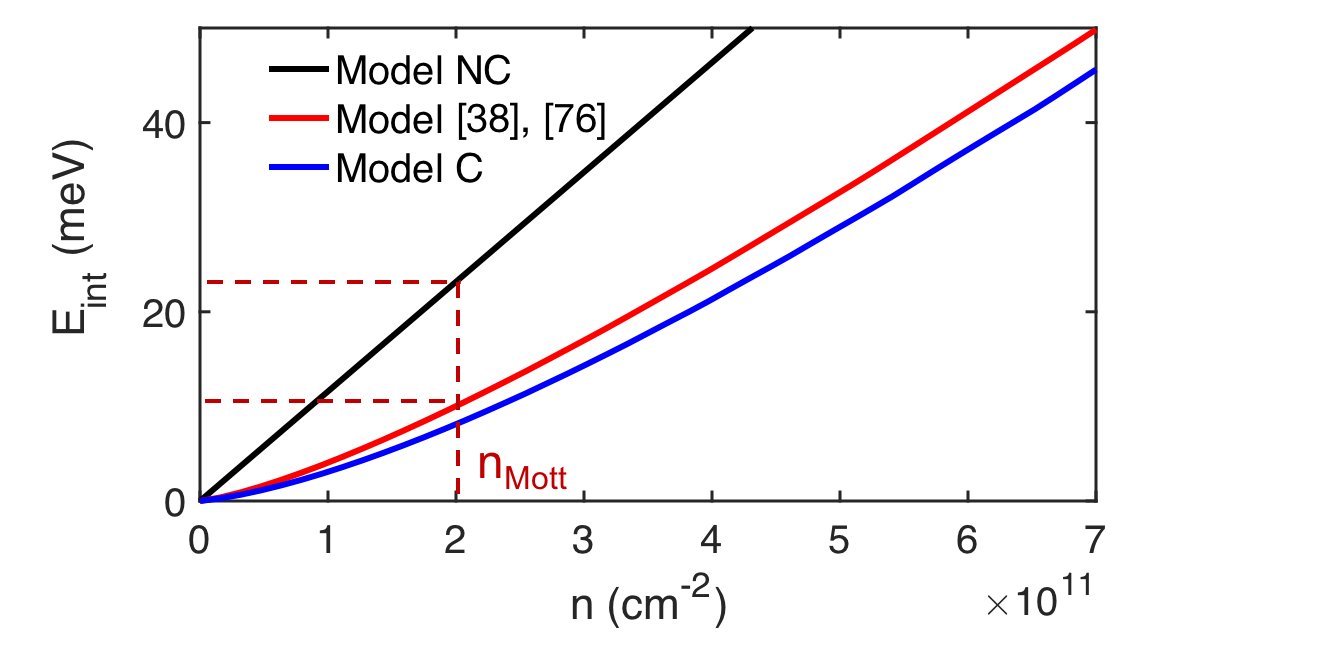}
    \caption{
    (a) Interaction energy (or, equivalently, density-induced energy shift) as a function of particle density for $T=$~7K. Model NC (self-consistent solution of the coupled Schr\"odinger and Poisson equations, Sec.~\ref{par:SP}), black line; Model C solved with (blue line) and without adaptive procedure, see Sec.\ref{par:model_C}. Dashed lines indicate the estimated Mott density and the corresponding interaction-induced energy shift.
}
    \label{fig:EintComputed}
\end{figure}

 According to~\cite{Zimmermann,ZimmermannComment2010}, this local quantity is given by the equation 
\begin{align}
\label{eq:nlZimmermann}
 1-e^ {-T_0(r)/T}=\left(1-e^ {-T_0^ {(0)}/T} \right) e^{-\left(U(r)+\phi_0(n_\ell-n)\right)/\left(k_B T \right)},
\end{align}
where
  $k_BT_0(r)=\pi\hbar^2 n_\ell(r,T,n)/2M$, represents the local chemical potential and
  $k_BT_0^ {(0)}=\pi\hbar^ 2n/(2M)$, represents chemical potential in the $r\to \infty$ limit.
The numerical resolution of this equation makes it possible to evaluate the local density $n_\ell(r,T,n)$ 
 We push the numerical  solution slightly further that the implementation presented in \cite{ZimmermannComment2010}. Our method is based on an adaptative calculation.  
 The solution $n_\ell(r,T,n)$  of Eq. \ref{eq:nlZimmermann}  is reinjected back in Eq.~\ref{eq:nl}. 
 The resulting $E_{\mathrm{int}}^\mathrm{corr}(n,T)$ is used to replace the "mean field" term $\phi_0(n_\ell-n)$ in Eq.~\ref{eq:nlZimmermann} by  $E_{\mathrm{int}}^\mathrm{corr}(n_\ell,T)-E_{\mathrm{int}}^\mathrm{corr}(n,T)$. 
 The procedure is repeated until convergence.

The density dependence of the pair correlation fonction $g(r)= n_\ell(r,T,n)/n$ obtained within this model is shown in Fig.~\ref{fig:paireCorrelationFonction}. Because of the indirect nature of the excitons, a depletion area appears around each exciton. 
This area is characterized by an exclusion radius $r_0(T,n)$.  One can clearly see in Fig.~\ref{fig:paireCorrelationFonction}, that the shape of the $g(r)$  gets steeper when the pair density increases.
At low density the exclusion radius is given by the mean field value 
\begin{align}
\label{eq:r0}
r_0(n=0,T)=\left(\frac{e^2d^2}{4 \pi \epsilon} \frac{1}{k_BT}\right)^{1/3},
\end{align}
that is  $\sim 30$~nm at $7$~K, and it shrinks down up to $\sim 5$~nm at highest densities.
Interaction energy $E_{\mathrm{int}}(n,T)$ calculated within Model C is shown in Fig.~\ref{fig:EintComputed}, for T=7~K. The red line corresponds to the implementation Refs.~\cite{Zimmermann,ZimmermannComment2010} and the blue line to the adaptive calculation described above and implemented  in the main text to estimate the densities.  
%
%

\bibliography{biblioGaN}

\begin{thebibliography}{76}%
\makeatletter
\providecommand \@ifxundefined [1]{%
 \@ifx{#1\undefined}
}%
\providecommand \@ifnum [1]{%
 \ifnum #1\expandafter \@firstoftwo
 \else \expandafter \@secondoftwo
 \fi
}%
\providecommand \@ifx [1]{%
 \ifx #1\expandafter \@firstoftwo
 \else \expandafter \@secondoftwo
 \fi
}%
\providecommand \natexlab [1]{#1}%
\providecommand \enquote  [1]{``#1''}%
\providecommand \bibnamefont  [1]{#1}%
\providecommand \bibfnamefont [1]{#1}%
\providecommand \citenamefont [1]{#1}%
\providecommand \href@noop [0]{\@secondoftwo}%
\providecommand \href [0]{\begingroup \@sanitize@url \@href}%
\providecommand \@href[1]{\@@startlink{#1}\@@href}%
\providecommand \@@href[1]{\endgroup#1\@@endlink}%
\providecommand \@sanitize@url [0]{\catcode `\\12\catcode `\$12\catcode
  `\&12\catcode `\#12\catcode `\^12\catcode `\_12\catcode `\%12\relax}%
\providecommand \@@startlink[1]{}%
\providecommand \@@endlink[0]{}%
\providecommand \url  [0]{\begingroup\@sanitize@url \@url }%
\providecommand \@url [1]{\endgroup\@href {#1}{\urlprefix }}%
\providecommand \urlprefix  [0]{URL }%
\providecommand \Eprint [0]{\href }%
\providecommand \doibase [0]{http://dx.doi.org/}%
\providecommand \selectlanguage [0]{\@gobble}%
\providecommand \bibinfo  [0]{\@secondoftwo}%
\providecommand \bibfield  [0]{\@secondoftwo}%
\providecommand \translation [1]{[#1]}%
\providecommand \BibitemOpen [0]{}%
\providecommand \bibitemStop [0]{}%
\providecommand \bibitemNoStop [0]{.\EOS\space}%
\providecommand \EOS [0]{\spacefactor3000\relax}%
\providecommand \BibitemShut  [1]{\csname bibitem#1\endcsname}%
\let\auto@bib@innerbib\@empty
\bibitem [{\citenamefont {Yudson}\ and\ \citenamefont
  {Lozovik}(1976)}]{LozovikYudson}%
  \BibitemOpen
  \bibfield  {author} {\bibinfo {author} {\bibfnamefont {V~I}\ \bibnamefont
  {Yudson}}\ and\ \bibinfo {author} {\bibfnamefont {Yu~E}\ \bibnamefont
  {Lozovik}},\ }\bibfield  {title} {\enquote {\bibinfo {title} {{A new
  mechanism for superconductivity: pairing between spatially separated
  electrons and holes}},}\ }\href@noop {} {\bibfield  {journal} {\bibinfo
  {journal} {Soviet Physics JETP}\ }\textbf {\bibinfo {volume} {44}},\ \bibinfo
  {pages} {389} (\bibinfo {year} {1976})}\BibitemShut {NoStop}%
\bibitem [{\citenamefont {Miller}\ \emph {et~al.}(1985)\citenamefont {Miller},
  \citenamefont {Chemla}, \citenamefont {Damen}, \citenamefont {Gossard},
  \citenamefont {Wiegmann}, \citenamefont {Wood},\ and\ \citenamefont
  {Burrus}}]{Miller1985}%
  \BibitemOpen
  \bibfield  {author} {\bibinfo {author} {\bibfnamefont {D~A~B}\ \bibnamefont
  {Miller}}, \bibinfo {author} {\bibfnamefont {D~S}\ \bibnamefont {Chemla}},
  \bibinfo {author} {\bibfnamefont {T~C}\ \bibnamefont {Damen}}, \bibinfo
  {author} {\bibfnamefont {A~C}\ \bibnamefont {Gossard}}, \bibinfo {author}
  {\bibfnamefont {W}~\bibnamefont {Wiegmann}}, \bibinfo {author} {\bibfnamefont
  {T~H}\ \bibnamefont {Wood}}, \ and\ \bibinfo {author} {\bibfnamefont {C~A}\
  \bibnamefont {Burrus}},\ }\bibfield  {title} {\enquote {\bibinfo {title}
  {{Electric-Field Dependence of Optical-Absorption Near the Band-Gap of
  Quantum-Well Structures}},}\ }\href@noop {} {\bibfield  {journal} {\bibinfo
  {journal} {Phys. Rev. B}\ }\textbf {\bibinfo {volume} {32}},\ \bibinfo
  {pages} {1043--1060} (\bibinfo {year} {1985})}\BibitemShut {NoStop}%
\bibitem [{\citenamefont {Huber}\ \emph {et~al.}(1998)\citenamefont {Huber},
  \citenamefont {Zrenner}, \citenamefont {Wegscheider},\ and\ \citenamefont
  {Bichler}}]{Huber1998}%
  \BibitemOpen
  \bibfield  {author} {\bibinfo {author} {\bibfnamefont {T}~\bibnamefont
  {Huber}}, \bibinfo {author} {\bibfnamefont {A}~\bibnamefont {Zrenner}},
  \bibinfo {author} {\bibfnamefont {W}~\bibnamefont {Wegscheider}}, \ and\
  \bibinfo {author} {\bibfnamefont {M}~\bibnamefont {Bichler}},\ }\bibfield
  {title} {\enquote {\bibinfo {title} {{Electrostatic Exciton Traps}},}\
  }\href@noop {} {\bibfield  {journal} {\bibinfo  {journal} {physica status
  solidi (a)}\ }\textbf {\bibinfo {volume} {166}},\ \bibinfo {pages} {R5--R6}
  (\bibinfo {year} {1998})}\BibitemShut {NoStop}%
\bibitem [{\citenamefont {Ivanov}\ \emph {et~al.}(1999)\citenamefont {Ivanov},
  \citenamefont {Littlewood},\ and\ \citenamefont {Haug}}]{Ivanov1999}%
  \BibitemOpen
  \bibfield  {author} {\bibinfo {author} {\bibfnamefont {A~L}\ \bibnamefont
  {Ivanov}}, \bibinfo {author} {\bibfnamefont {P~B}\ \bibnamefont
  {Littlewood}}, \ and\ \bibinfo {author} {\bibfnamefont {H}~\bibnamefont
  {Haug}},\ }\bibfield  {title} {\enquote {\bibinfo {title} {{Bose-Einstein
  statistics in thermalization and photoluminescence of quantum-well
  excitons}},}\ }\href@noop {} {\bibfield  {journal} {\bibinfo  {journal}
  {Phys. Rev. B}\ }\textbf {\bibinfo {volume} {59}},\ \bibinfo {pages}
  {5032--5048} (\bibinfo {year} {1999})}\BibitemShut {NoStop}%
\bibitem [{\citenamefont {Butov}\ \emph {et~al.}(1999)\citenamefont {Butov},
  \citenamefont {Shashkin}, \citenamefont {Dolgopolov}, \citenamefont
  {Campman},\ and\ \citenamefont {Gossard}}]{Butov1999}%
  \BibitemOpen
  \bibfield  {author} {\bibinfo {author} {\bibfnamefont {L~V}\ \bibnamefont
  {Butov}}, \bibinfo {author} {\bibfnamefont {A~A}\ \bibnamefont {Shashkin}},
  \bibinfo {author} {\bibfnamefont {V~T}\ \bibnamefont {Dolgopolov}}, \bibinfo
  {author} {\bibfnamefont {K~L}\ \bibnamefont {Campman}}, \ and\ \bibinfo
  {author} {\bibfnamefont {A~C}\ \bibnamefont {Gossard}},\ }\bibfield  {title}
  {\enquote {\bibinfo {title} {{Magneto-optics of the spatially separated
  electron and hole layers in GaAs/AlxGa1-xAs coupled quantum wells}},}\
  }\href@noop {} {\bibfield  {journal} {\bibinfo  {journal} {Phys. Rev. B}\
  }\textbf {\bibinfo {volume} {60}},\ \bibinfo {pages} {8753--8758} (\bibinfo
  {year} {1999})}\BibitemShut {NoStop}%
\bibitem [{\citenamefont {High}\ \emph {et~al.}(2008)\citenamefont {High},
  \citenamefont {Novitskaya}, \citenamefont {Butov}, \citenamefont {Hanson},\
  and\ \citenamefont {Gossard}}]{High2008}%
  \BibitemOpen
  \bibfield  {author} {\bibinfo {author} {\bibfnamefont {Alex~A}\ \bibnamefont
  {High}}, \bibinfo {author} {\bibfnamefont {Ekaterina~E}\ \bibnamefont
  {Novitskaya}}, \bibinfo {author} {\bibfnamefont {Leonid~V}\ \bibnamefont
  {Butov}}, \bibinfo {author} {\bibfnamefont {Micah}\ \bibnamefont {Hanson}}, \
  and\ \bibinfo {author} {\bibfnamefont {Arthur~C}\ \bibnamefont {Gossard}},\
  }\bibfield  {title} {\enquote {\bibinfo {title} {{Control of exciton fluxes
  in an excitonic integrated circuit.}}}\ }\href@noop {} {\bibfield  {journal}
  {\bibinfo  {journal} {Science}\ }\textbf {\bibinfo {volume} {321}},\ \bibinfo
  {pages} {229--231} (\bibinfo {year} {2008})}\BibitemShut {NoStop}%
\bibitem [{\citenamefont {Butov}(2017)}]{Butov2017}%
  \BibitemOpen
  \bibfield  {author} {\bibinfo {author} {\bibfnamefont {L~V}\ \bibnamefont
  {Butov}},\ }\bibfield  {title} {\enquote {\bibinfo {title} {{Excitonic
  devices}},}\ }\href@noop {} {\bibfield  {journal} {\bibinfo  {journal}
  {Superlattices and Microstructures}\ }\textbf {\bibinfo {volume} {108}},\
  \bibinfo {pages} {2--26} (\bibinfo {year} {2017})}\BibitemShut {NoStop}%
\bibitem [{\citenamefont {Lozovik}\ and\ \citenamefont
  {Berman}(1996)}]{Lozovik1996}%
  \BibitemOpen
  \bibfield  {author} {\bibinfo {author} {\bibfnamefont {Yu~E}\ \bibnamefont
  {Lozovik}}\ and\ \bibinfo {author} {\bibfnamefont {O~L}\ \bibnamefont
  {Berman}},\ }\bibfield  {title} {\enquote {\bibinfo {title} {{Phase
  transitions in a system of two coupled quantum wells}},}\ }\href@noop {}
  {\bibfield  {journal} {\bibinfo  {journal} {Jetp Lett.}\ }\textbf {\bibinfo
  {volume} {64}},\ \bibinfo {pages} {573--579} (\bibinfo {year}
  {1996})}\BibitemShut {NoStop}%
\bibitem [{\citenamefont {High}\ \emph {et~al.}(2012)\citenamefont {High},
  \citenamefont {Leonard}, \citenamefont {Hammack}, \citenamefont {Fogler},
  \citenamefont {Butov}, \citenamefont {Kavokin}, \citenamefont {Campman},\
  and\ \citenamefont {Gossard}}]{High2012}%
  \BibitemOpen
  \bibfield  {author} {\bibinfo {author} {\bibfnamefont {A~A}\ \bibnamefont
  {High}}, \bibinfo {author} {\bibfnamefont {J~R}\ \bibnamefont {Leonard}},
  \bibinfo {author} {\bibfnamefont {A~T}\ \bibnamefont {Hammack}}, \bibinfo
  {author} {\bibfnamefont {M~M}\ \bibnamefont {Fogler}}, \bibinfo {author}
  {\bibfnamefont {L~V}\ \bibnamefont {Butov}}, \bibinfo {author} {\bibfnamefont
  {A~V}\ \bibnamefont {Kavokin}}, \bibinfo {author} {\bibfnamefont {K~L}\
  \bibnamefont {Campman}}, \ and\ \bibinfo {author} {\bibfnamefont {A~C}\
  \bibnamefont {Gossard}},\ }\bibfield  {title} {\enquote {\bibinfo {title}
  {{Spontaneous coherence in a cold exciton gas}},}\ }\href@noop {} {\bibfield
  {journal} {\bibinfo  {journal} {Nature}\ }\textbf {\bibinfo {volume} {483}},\
  \bibinfo {pages} {584--588} (\bibinfo {year} {2012})}\BibitemShut {NoStop}%
\bibitem [{\citenamefont {Shilo}\ \emph {et~al.}(2013)\citenamefont {Shilo},
  \citenamefont {Cohen}, \citenamefont {Laikhtman}, \citenamefont {West},
  \citenamefont {Pfeiffer},\ and\ \citenamefont {Rapaport}}]{Shilo2013}%
  \BibitemOpen
  \bibfield  {author} {\bibinfo {author} {\bibfnamefont {Yehiel}\ \bibnamefont
  {Shilo}}, \bibinfo {author} {\bibfnamefont {Kobi}\ \bibnamefont {Cohen}},
  \bibinfo {author} {\bibfnamefont {Boris}\ \bibnamefont {Laikhtman}}, \bibinfo
  {author} {\bibfnamefont {Ken}\ \bibnamefont {West}}, \bibinfo {author}
  {\bibfnamefont {Loren}\ \bibnamefont {Pfeiffer}}, \ and\ \bibinfo {author}
  {\bibfnamefont {Ronen}\ \bibnamefont {Rapaport}},\ }\bibfield  {title}
  {\enquote {\bibinfo {title} {{Particle correlations and evidence for dark
  state condensation in a cold dipolar exciton fluid}},}\ }\href@noop {}
  {\bibfield  {journal} {\bibinfo  {journal} {Nature Communications}\ }\textbf
  {\bibinfo {volume} {4}},\ \bibinfo {pages} {1--7} (\bibinfo {year}
  {2013})}\BibitemShut {NoStop}%
\bibitem [{\citenamefont {Schinner}\ \emph {et~al.}(2013)\citenamefont
  {Schinner}, \citenamefont {Repp}, \citenamefont {Schubert}, \citenamefont
  {Rai}, \citenamefont {Reuter}, \citenamefont {Wieck}, \citenamefont
  {Govorov}, \citenamefont {Holleitner},\ and\ \citenamefont
  {Kotthaus}}]{Schinner2013}%
  \BibitemOpen
  \bibfield  {author} {\bibinfo {author} {\bibfnamefont {G~J}\ \bibnamefont
  {Schinner}}, \bibinfo {author} {\bibfnamefont {J}~\bibnamefont {Repp}},
  \bibinfo {author} {\bibfnamefont {E}~\bibnamefont {Schubert}}, \bibinfo
  {author} {\bibfnamefont {A~K}\ \bibnamefont {Rai}}, \bibinfo {author}
  {\bibfnamefont {D}~\bibnamefont {Reuter}}, \bibinfo {author} {\bibfnamefont
  {A~D}\ \bibnamefont {Wieck}}, \bibinfo {author} {\bibfnamefont {A~O}\
  \bibnamefont {Govorov}}, \bibinfo {author} {\bibfnamefont {A~W}\ \bibnamefont
  {Holleitner}}, \ and\ \bibinfo {author} {\bibfnamefont {J~P}\ \bibnamefont
  {Kotthaus}},\ }\bibfield  {title} {\enquote {\bibinfo {title} {{Many-body
  correlations of electrostatically trapped dipolar excitons}},}\ }\href@noop
  {} {\bibfield  {journal} {\bibinfo  {journal} {Phys. Rev. B}\ }\textbf
  {\bibinfo {volume} {87}},\ \bibinfo {pages} {205302} (\bibinfo {year}
  {2013})}\BibitemShut {NoStop}%
\bibitem [{\citenamefont {Cohen}\ \emph {et~al.}(2016)\citenamefont {Cohen},
  \citenamefont {Shilo}, \citenamefont {West}, \citenamefont {Pfeiffer},\ and\
  \citenamefont {Rapaport}}]{Cohen2016}%
  \BibitemOpen
  \bibfield  {author} {\bibinfo {author} {\bibfnamefont {Kobi}\ \bibnamefont
  {Cohen}}, \bibinfo {author} {\bibfnamefont {Yehiel}\ \bibnamefont {Shilo}},
  \bibinfo {author} {\bibfnamefont {Ken}\ \bibnamefont {West}}, \bibinfo
  {author} {\bibfnamefont {Loren}\ \bibnamefont {Pfeiffer}}, \ and\ \bibinfo
  {author} {\bibfnamefont {Ronen}\ \bibnamefont {Rapaport}},\ }\bibfield
  {title} {\enquote {\bibinfo {title} {{Dark High Density Dipolar Liquid of
  Excitons}},}\ }\href@noop {} {\bibfield  {journal} {\bibinfo  {journal} {Nano
  Lett.}\ }\textbf {\bibinfo {volume} {16}},\ \bibinfo {pages} {3726--3731}
  (\bibinfo {year} {2016})}\BibitemShut {NoStop}%
\bibitem [{\citenamefont {Butov}(2016)}]{Butov2016}%
  \BibitemOpen
  \bibfield  {author} {\bibinfo {author} {\bibfnamefont {L~V}\ \bibnamefont
  {Butov}},\ }\bibfield  {title} {\enquote {\bibinfo {title} {{Collective
  phenomena in cold indirect excitons}},}\ }\href@noop {} {\bibfield  {journal}
  {\bibinfo  {journal} {J. Exp. Theor. Phys.}\ }\textbf {\bibinfo {volume}
  {149}},\ \bibinfo {pages} {505} (\bibinfo {year} {2016})}\BibitemShut
  {NoStop}%
\bibitem [{\citenamefont {Combescot}\ \emph {et~al.}(2017)\citenamefont
  {Combescot}, \citenamefont {Combescot},\ and\ \citenamefont
  {Dubin}}]{Combescot2017}%
  \BibitemOpen
  \bibfield  {author} {\bibinfo {author} {\bibfnamefont {Monique}\ \bibnamefont
  {Combescot}}, \bibinfo {author} {\bibfnamefont {Roland}\ \bibnamefont
  {Combescot}}, \ and\ \bibinfo {author} {\bibfnamefont {Fran{\c c}ois}\
  \bibnamefont {Dubin}},\ }\bibfield  {title} {\enquote {\bibinfo {title}
  {{Bose{\textendash}Einstein condensation and indirect excitons: a review}},}\
  }\href@noop {} {\bibfield  {journal} {\bibinfo  {journal} {Rep. Prog. Phys.}\
  }\textbf {\bibinfo {volume} {80}},\ \bibinfo {pages} {066501} (\bibinfo
  {year} {2017})}\BibitemShut {NoStop}%
\bibitem [{\citenamefont {Anankine}\ \emph {et~al.}(2017)\citenamefont
  {Anankine}, \citenamefont {Beian}, \citenamefont {Dang}, \citenamefont
  {Alloing}, \citenamefont {Cambril}, \citenamefont {Merghem}, \citenamefont
  {Carbonell}, \citenamefont {Lema{\^\i}tre},\ and\ \citenamefont
  {Dubin}}]{Anankine2017}%
  \BibitemOpen
  \bibfield  {author} {\bibinfo {author} {\bibfnamefont {Romain}\ \bibnamefont
  {Anankine}}, \bibinfo {author} {\bibfnamefont {Mussie}\ \bibnamefont
  {Beian}}, \bibinfo {author} {\bibfnamefont {Suzanne}\ \bibnamefont {Dang}},
  \bibinfo {author} {\bibfnamefont {Mathieu}\ \bibnamefont {Alloing}}, \bibinfo
  {author} {\bibfnamefont {Edmond}\ \bibnamefont {Cambril}}, \bibinfo {author}
  {\bibfnamefont {Kamel}\ \bibnamefont {Merghem}}, \bibinfo {author}
  {\bibfnamefont {Carmen~Gomez}\ \bibnamefont {Carbonell}}, \bibinfo {author}
  {\bibfnamefont {Aristide}\ \bibnamefont {Lema{\^\i}tre}}, \ and\ \bibinfo
  {author} {\bibfnamefont {Fran{\c c}ois}\ \bibnamefont {Dubin}},\ }\bibfield
  {title} {\enquote {\bibinfo {title} {{Quantized Vortices and Four-Component
  Superfluidity of Semiconductor Excitons}},}\ }\href@noop {} {\bibfield
  {journal} {\bibinfo  {journal} {Phys. Rev. Lett.}\ }\textbf {\bibinfo
  {volume} {118}},\ \bibinfo {pages} {127402} (\bibinfo {year}
  {2017})}\BibitemShut {NoStop}%
\bibitem [{\citenamefont {Misra}\ \emph {et~al.}(2018)\citenamefont {Misra},
  \citenamefont {Stern}, \citenamefont {Joshua}, \citenamefont {Umansky},\ and\
  \citenamefont {Bar-Joseph}}]{Misra2018}%
  \BibitemOpen
  \bibfield  {author} {\bibinfo {author} {\bibfnamefont {Subhradeep}\
  \bibnamefont {Misra}}, \bibinfo {author} {\bibfnamefont {Michael}\
  \bibnamefont {Stern}}, \bibinfo {author} {\bibfnamefont {Arjun}\ \bibnamefont
  {Joshua}}, \bibinfo {author} {\bibfnamefont {Vladimir}\ \bibnamefont
  {Umansky}}, \ and\ \bibinfo {author} {\bibfnamefont {Israel}\ \bibnamefont
  {Bar-Joseph}},\ }\bibfield  {title} {\enquote {\bibinfo {title}
  {{Experimental Study of the Exciton Gas-Liquid Transition in Coupled Quantum
  Wells}},}\ }\href@noop {} {\bibfield  {journal} {\bibinfo  {journal} {Phys.
  Rev. Lett.}\ }\textbf {\bibinfo {volume} {120}},\ \bibinfo {pages} {047402}
  (\bibinfo {year} {2018})}\BibitemShut {NoStop}%
\bibitem [{\citenamefont {Laikhtman}\ and\ \citenamefont
  {Rapaport}(2009{\natexlab{a}})}]{Laikhtman2009}%
  \BibitemOpen
  \bibfield  {author} {\bibinfo {author} {\bibfnamefont {B}~\bibnamefont
  {Laikhtman}}\ and\ \bibinfo {author} {\bibfnamefont {Ronen}\ \bibnamefont
  {Rapaport}},\ }\bibfield  {title} {\enquote {\bibinfo {title} {{Exciton
  correlations in coupled quantum wells and their luminescence blue shift}},}\
  }\href@noop {} {\bibfield  {journal} {\bibinfo  {journal} {Phys. Rev. B}\
  }\textbf {\bibinfo {volume} {80}},\ \bibinfo {pages} {195313--12} (\bibinfo
  {year} {2009}{\natexlab{a}})}\BibitemShut {NoStop}%
\bibitem [{\citenamefont {Zimmermann}\ \emph {et~al.}(1978)\citenamefont
  {Zimmermann}, \citenamefont {Kilimann}, \citenamefont {Kraeft}, \citenamefont
  {Kremp},\ and\ \citenamefont {R{\"o}pke}}]{Zimmermann1978}%
  \BibitemOpen
  \bibfield  {author} {\bibinfo {author} {\bibfnamefont {R}~\bibnamefont
  {Zimmermann}}, \bibinfo {author} {\bibfnamefont {K}~\bibnamefont {Kilimann}},
  \bibinfo {author} {\bibfnamefont {W~D}\ \bibnamefont {Kraeft}}, \bibinfo
  {author} {\bibfnamefont {D}~\bibnamefont {Kremp}}, \ and\ \bibinfo {author}
  {\bibfnamefont {G}~\bibnamefont {R{\"o}pke}},\ }\bibfield  {title} {\enquote
  {\bibinfo {title} {{Dynamical screening and self-energy of excitons in the
  electron{\textendash}hole plasma}},}\ }\href@noop {} {\bibfield  {journal}
  {\bibinfo  {journal} {phys. stat. sol. (b)}\ }\textbf {\bibinfo {volume}
  {90}},\ \bibinfo {pages} {175--187} (\bibinfo {year} {1978})}\BibitemShut
  {NoStop}%
\bibitem [{\citenamefont {Mott}(1968)}]{Mott1968}%
  \BibitemOpen
  \bibfield  {author} {\bibinfo {author} {\bibfnamefont {N~F}\ \bibnamefont
  {Mott}},\ }\bibfield  {title} {\enquote {\bibinfo {title} {{Metal-Insulator
  Transition}},}\ }\href@noop {} {\bibfield  {journal} {\bibinfo  {journal}
  {Rev. Mod. Phys.}\ }\textbf {\bibinfo {volume} {40}},\ \bibinfo {pages}
  {677--683} (\bibinfo {year} {1968})}\BibitemShut {NoStop}%
\bibitem [{\citenamefont {Ben-Tabou~de Leon}\ and\ \citenamefont
  {Laikhtman}(2003)}]{BenTaboudeLeon2003}%
  \BibitemOpen
  \bibfield  {author} {\bibinfo {author} {\bibfnamefont {S}~\bibnamefont
  {Ben-Tabou~de Leon}}\ and\ \bibinfo {author} {\bibfnamefont {B}~\bibnamefont
  {Laikhtman}},\ }\bibfield  {title} {\enquote {\bibinfo {title} {{Mott
  transition, biexciton crossover, and spin ordering in the exciton gas in
  quantum wells}},}\ }\href@noop {} {\bibfield  {journal} {\bibinfo  {journal}
  {Phys. Rev. B}\ }\textbf {\bibinfo {volume} {67}},\ \bibinfo {pages} {235315}
  (\bibinfo {year} {2003})}\BibitemShut {NoStop}%
\bibitem [{\citenamefont {Nikolaev}\ and\ \citenamefont
  {Portnoi}(2004)}]{Nikolaev2004}%
  \BibitemOpen
  \bibfield  {author} {\bibinfo {author} {\bibfnamefont {V~V}\ \bibnamefont
  {Nikolaev}}\ and\ \bibinfo {author} {\bibfnamefont {M~E}\ \bibnamefont
  {Portnoi}},\ }\bibfield  {title} {\enquote {\bibinfo {title} {{Theory of
  excitonic Mott transition in double quantum wells}},}\ }\href@noop {}
  {\bibfield  {journal} {\bibinfo  {journal} {phys. stat. sol. (c)}\ }\textbf
  {\bibinfo {volume} {1}},\ \bibinfo {pages} {1357--1362} (\bibinfo {year}
  {2004})}\BibitemShut {NoStop}%
\bibitem [{\citenamefont {Stern}\ \emph {et~al.}(2008)\citenamefont {Stern},
  \citenamefont {Garmider}, \citenamefont {Umansky},\ and\ \citenamefont
  {Bar-Joseph}}]{Stern2008}%
  \BibitemOpen
  \bibfield  {author} {\bibinfo {author} {\bibfnamefont {M}~\bibnamefont
  {Stern}}, \bibinfo {author} {\bibfnamefont {V}~\bibnamefont {Garmider}},
  \bibinfo {author} {\bibfnamefont {V}~\bibnamefont {Umansky}}, \ and\ \bibinfo
  {author} {\bibfnamefont {I}~\bibnamefont {Bar-Joseph}},\ }\bibfield  {title}
  {\enquote {\bibinfo {title} {{Mott Transition of Excitons in Coupled Quantum
  Wells}},}\ }\href@noop {} {\bibfield  {journal} {\bibinfo  {journal} {Phys.
  Rev. Lett.}\ }\textbf {\bibinfo {volume} {100}},\ \bibinfo {pages} {256402}
  (\bibinfo {year} {2008})}\BibitemShut {NoStop}%
\bibitem [{\citenamefont {Snoke}(2008)}]{Snoke2008}%
  \BibitemOpen
  \bibfield  {author} {\bibinfo {author} {\bibfnamefont {David}\ \bibnamefont
  {Snoke}},\ }\bibfield  {title} {\enquote {\bibinfo {title} {{Predicting the
  ionization threshold for carriers in excited semiconductors}},}\ }\href@noop
  {} {\bibfield  {journal} {\bibinfo  {journal} {Solid State Communications}\
  }\textbf {\bibinfo {volume} {146}},\ \bibinfo {pages} {73} (\bibinfo {year}
  {2008})}\BibitemShut {NoStop}%
\bibitem [{\citenamefont {Byrnes}\ \emph {et~al.}(2010)\citenamefont {Byrnes},
  \citenamefont {Recher},\ and\ \citenamefont {Yamamoto}}]{Byrnes2010}%
  \BibitemOpen
  \bibfield  {author} {\bibinfo {author} {\bibfnamefont {Tim}\ \bibnamefont
  {Byrnes}}, \bibinfo {author} {\bibfnamefont {Patrik}\ \bibnamefont {Recher}},
  \ and\ \bibinfo {author} {\bibfnamefont {Yoshihisa}\ \bibnamefont
  {Yamamoto}},\ }\bibfield  {title} {\enquote {\bibinfo {title} {{Mott
  transitions of exciton polaritons and indirect excitons in a periodic
  potential}},}\ }\href@noop {} {\bibfield  {journal} {\bibinfo  {journal}
  {Phys. Rev. B}\ }\textbf {\bibinfo {volume} {81}},\ \bibinfo {pages}
  {205312--13} (\bibinfo {year} {2010})}\BibitemShut {NoStop}%
\bibitem [{\citenamefont {Kir{\v s}ansk{\.{e}}}\ \emph
  {et~al.}(2016)\citenamefont {Kir{\v s}ansk{\.{e}}}, \citenamefont
  {Tighineanu}, \citenamefont {Daveau}, \citenamefont {Miguel-S{\'a}nchez},
  \citenamefont {Lodahl},\ and\ \citenamefont {Stobbe}}]{Kirsanske2016}%
  \BibitemOpen
  \bibfield  {author} {\bibinfo {author} {\bibfnamefont {Gabija}\ \bibnamefont
  {Kir{\v s}ansk{\.{e}}}}, \bibinfo {author} {\bibfnamefont {Petru}\
  \bibnamefont {Tighineanu}}, \bibinfo {author} {\bibfnamefont {Rapha{\"e}l~S}\
  \bibnamefont {Daveau}}, \bibinfo {author} {\bibfnamefont {Javier}\
  \bibnamefont {Miguel-S{\'a}nchez}}, \bibinfo {author} {\bibfnamefont {Peter}\
  \bibnamefont {Lodahl}}, \ and\ \bibinfo {author} {\bibfnamefont {S{\o}ren}\
  \bibnamefont {Stobbe}},\ }\bibfield  {title} {\enquote {\bibinfo {title}
  {{Observation of the exciton Mott transition in the photoluminescence of
  coupled quantum wells}},}\ }\href@noop {} {\bibfield  {journal} {\bibinfo
  {journal} {Phys. Rev. B}\ }\textbf {\bibinfo {volume} {94}},\ \bibinfo
  {pages} {155438} (\bibinfo {year} {2016})}\BibitemShut {NoStop}%
\bibitem [{\citenamefont {Vignesh}\ and\ \citenamefont
  {Nithiananthi}(2020)}]{Vignesh2020}%
  \BibitemOpen
  \bibfield  {author} {\bibinfo {author} {\bibfnamefont {G}~\bibnamefont
  {Vignesh}}\ and\ \bibinfo {author} {\bibfnamefont {P}~\bibnamefont
  {Nithiananthi}},\ }\bibfield  {title} {\enquote {\bibinfo {title}
  {{Hartree-Fock approximation for exciton Mott transition in double quantum
  well: Direct and indirect exciton diamagnetism}},}\ }\href@noop {} {\bibfield
   {journal} {\bibinfo  {journal} {Physica E: Low-dimensional Systems and
  Nanostructures}\ ,\ \bibinfo {pages} {114008}} (\bibinfo {year}
  {2020})}\BibitemShut {NoStop}%
\bibitem [{\citenamefont {Kappei}\ \emph {et~al.}(2005)\citenamefont {Kappei},
  \citenamefont {Szczytko}, \citenamefont {Morier-Genoud},\ and\ \citenamefont
  {Deveaud}}]{Kappei2005}%
  \BibitemOpen
  \bibfield  {author} {\bibinfo {author} {\bibfnamefont {L}~\bibnamefont
  {Kappei}}, \bibinfo {author} {\bibfnamefont {J}~\bibnamefont {Szczytko}},
  \bibinfo {author} {\bibfnamefont {F}~\bibnamefont {Morier-Genoud}}, \ and\
  \bibinfo {author} {\bibfnamefont {B}~\bibnamefont {Deveaud}},\ }\bibfield
  {title} {\enquote {\bibinfo {title} {{Direct Observation of the Mott
  Transition in an Optically Excited Semiconductor Quantum Well}},}\
  }\href@noop {} {\bibfield  {journal} {\bibinfo  {journal} {Phys. Rev. Lett.}\
  }\textbf {\bibinfo {volume} {94}},\ \bibinfo {pages} {147403} (\bibinfo
  {year} {2005})}\BibitemShut {NoStop}%
\bibitem [{\citenamefont {Deveaud}\ \emph {et~al.}(2005)\citenamefont
  {Deveaud}, \citenamefont {Kappei}, \citenamefont {Berney}, \citenamefont
  {Morier-Genoud}, \citenamefont {Portella-Oberli}, \citenamefont {Szczytko},\
  and\ \citenamefont {Piermarocchi}}]{Deveaud2005}%
  \BibitemOpen
  \bibfield  {author} {\bibinfo {author} {\bibfnamefont {B}~\bibnamefont
  {Deveaud}}, \bibinfo {author} {\bibfnamefont {L}~\bibnamefont {Kappei}},
  \bibinfo {author} {\bibfnamefont {J}~\bibnamefont {Berney}}, \bibinfo
  {author} {\bibfnamefont {F}~\bibnamefont {Morier-Genoud}}, \bibinfo {author}
  {\bibfnamefont {M~T}\ \bibnamefont {Portella-Oberli}}, \bibinfo {author}
  {\bibfnamefont {J}~\bibnamefont {Szczytko}}, \ and\ \bibinfo {author}
  {\bibfnamefont {C}~\bibnamefont {Piermarocchi}},\ }\bibfield  {title}
  {\enquote {\bibinfo {title} {{Excitonic effects in the luminescence of
  quantum wells}},}\ }\href@noop {} {\bibfield  {journal} {\bibinfo  {journal}
  {Chemical Physics}\ }\textbf {\bibinfo {volume} {318}},\ \bibinfo {pages}
  {104--117} (\bibinfo {year} {2005})}\BibitemShut {NoStop}%
\bibitem [{\citenamefont {Nikolaev}\ and\ \citenamefont
  {Portnoi}(2008)}]{Nikolaev2008}%
  \BibitemOpen
  \bibfield  {author} {\bibinfo {author} {\bibfnamefont {V~V}\ \bibnamefont
  {Nikolaev}}\ and\ \bibinfo {author} {\bibfnamefont {M~E}\ \bibnamefont
  {Portnoi}},\ }\bibfield  {title} {\enquote {\bibinfo {title} {{Theory of the
  excitonic Mott transition in quasi-two-dimensional systems}},}\ }\href@noop
  {} {\bibfield  {journal} {\bibinfo  {journal} {Superlattices and
  Microstructures}\ }\textbf {\bibinfo {volume} {43}},\ \bibinfo {pages}
  {460--464} (\bibinfo {year} {2008})}\BibitemShut {NoStop}%
\bibitem [{\citenamefont {Sekiguchi}\ and\ \citenamefont
  {Shimano}(2017)}]{Sekiguchi2017}%
  \BibitemOpen
  \bibfield  {author} {\bibinfo {author} {\bibfnamefont {Fumiya}\ \bibnamefont
  {Sekiguchi}}\ and\ \bibinfo {author} {\bibfnamefont {Ryo}\ \bibnamefont
  {Shimano}},\ }\bibfield  {title} {\enquote {\bibinfo {title} {{Rate Equation
  Analysis of the Dynamics of First-order Exciton Mott Transition}},}\
  }\href@noop {} {\bibfield  {journal} {\bibinfo  {journal} {Journal of the
  Physical Society of Japan}\ }\textbf {\bibinfo {volume} {86}},\ \bibinfo
  {pages} {103702} (\bibinfo {year} {2017})}\BibitemShut {NoStop}%
\bibitem [{\citenamefont {Mock}\ \emph {et~al.}(1978)\citenamefont {Mock},
  \citenamefont {Thomas},\ and\ \citenamefont {Combescot}}]{Mock1978}%
  \BibitemOpen
  \bibfield  {author} {\bibinfo {author} {\bibfnamefont {J~B}\ \bibnamefont
  {Mock}}, \bibinfo {author} {\bibfnamefont {G~A}\ \bibnamefont {Thomas}}, \
  and\ \bibinfo {author} {\bibfnamefont {M}~\bibnamefont {Combescot}},\
  }\bibfield  {title} {\enquote {\bibinfo {title} {{Entropy ionization of an
  exciton gas}},}\ }\href@noop {} {\bibfield  {journal} {\bibinfo  {journal}
  {Solid State Communications}\ }\textbf {\bibinfo {volume} {25}},\ \bibinfo
  {pages} {279--282} (\bibinfo {year} {1978})}\BibitemShut {NoStop}%
\bibitem [{\citenamefont {Chiaruttini}\ \emph {et~al.}(2019)\citenamefont
  {Chiaruttini}, \citenamefont {Guillet}, \citenamefont {Brimont},
  \citenamefont {Jouault}, \citenamefont {Lefebvre}, \citenamefont {Vives},
  \citenamefont {Chenot}, \citenamefont {Cordier}, \citenamefont {Damilano},\
  and\ \citenamefont {Vladimirova}}]{Chiaruttini2019}%
  \BibitemOpen
  \bibfield  {author} {\bibinfo {author} {\bibfnamefont {Fran{\c c}ois}\
  \bibnamefont {Chiaruttini}}, \bibinfo {author} {\bibfnamefont {Thierry}\
  \bibnamefont {Guillet}}, \bibinfo {author} {\bibfnamefont {Christelle}\
  \bibnamefont {Brimont}}, \bibinfo {author} {\bibfnamefont {Benoit}\
  \bibnamefont {Jouault}}, \bibinfo {author} {\bibfnamefont {Pierre}\
  \bibnamefont {Lefebvre}}, \bibinfo {author} {\bibfnamefont {Jessica}\
  \bibnamefont {Vives}}, \bibinfo {author} {\bibfnamefont {Sebastien}\
  \bibnamefont {Chenot}}, \bibinfo {author} {\bibfnamefont {Yvon}\ \bibnamefont
  {Cordier}}, \bibinfo {author} {\bibfnamefont {Benjamin}\ \bibnamefont
  {Damilano}}, \ and\ \bibinfo {author} {\bibfnamefont {Maria}\ \bibnamefont
  {Vladimirova}},\ }\bibfield  {title} {\enquote {\bibinfo {title} {{Trapping
  Dipolar Exciton Fluids in GaN/(AlGa)N Nanostructures}},}\ }\href@noop {}
  {\bibfield  {journal} {\bibinfo  {journal} {Nano Lett.}\ }\textbf {\bibinfo
  {volume} {19}},\ \bibinfo {pages} {4911} (\bibinfo {year}
  {2019})}\BibitemShut {NoStop}%
\bibitem [{\citenamefont {Gil}(2014)}]{Gil2014}%
  \BibitemOpen
  \bibfield  {author} {\bibinfo {author} {\bibfnamefont {Bernard}\ \bibnamefont
  {Gil}},\ }\href@noop {} {\emph {\bibinfo {title} {III-Nitride Semiconductors
  and their Modern Devices}}}\ (\bibinfo  {publisher} {Springer},\ \bibinfo
  {year} {2014})\BibitemShut {NoStop}%
\bibitem [{\citenamefont {Bernardini}\ and\ \citenamefont
  {Fiorentini}(1998)}]{Bernardini1997}%
  \BibitemOpen
  \bibfield  {author} {\bibinfo {author} {\bibfnamefont {Fabio}\ \bibnamefont
  {Bernardini}}\ and\ \bibinfo {author} {\bibfnamefont {Vincenzo}\ \bibnamefont
  {Fiorentini}},\ }\bibfield  {title} {\enquote {\bibinfo {title} {{Macroscopic
  polarization and band offsets at nitride heterojunctions}},}\ }\href@noop {}
  {\bibfield  {journal} {\bibinfo  {journal} {Phys. Rev. B}\ }\textbf {\bibinfo
  {volume} {57}},\ \bibinfo {pages} {R9427--R9430} (\bibinfo {year}
  {1998})}\BibitemShut {NoStop}%
\bibitem [{\citenamefont {Leroux}\ \emph {et~al.}(1998)\citenamefont {Leroux},
  \citenamefont {Grandjean}, \citenamefont {La{\"u}gt}, \citenamefont
  {Massies}, \citenamefont {Gil}, \citenamefont {Lefebvre},\ and\ \citenamefont
  {Bigenwald}}]{Leroux1998}%
  \BibitemOpen
  \bibfield  {author} {\bibinfo {author} {\bibfnamefont {M}~\bibnamefont
  {Leroux}}, \bibinfo {author} {\bibfnamefont {N}~\bibnamefont {Grandjean}},
  \bibinfo {author} {\bibfnamefont {M}~\bibnamefont {La{\"u}gt}}, \bibinfo
  {author} {\bibfnamefont {J}~\bibnamefont {Massies}}, \bibinfo {author}
  {\bibfnamefont {B}~\bibnamefont {Gil}}, \bibinfo {author} {\bibfnamefont
  {P}~\bibnamefont {Lefebvre}}, \ and\ \bibinfo {author} {\bibfnamefont
  {P}~\bibnamefont {Bigenwald}},\ }\bibfield  {title} {\enquote {\bibinfo
  {title} {{Quantum confined Stark effect due to built-in internal polarization
  fields in (Al,Ga)N/GaN quantum wells}},}\ }\href@noop {} {\bibfield
  {journal} {\bibinfo  {journal} {Phys. Rev. B}\ }\textbf {\bibinfo {volume}
  {58}},\ \bibinfo {pages} {R13371--R13374} (\bibinfo {year}
  {1998})}\BibitemShut {NoStop}%
\bibitem [{\citenamefont {Grandjean}\ \emph {et~al.}(1999)\citenamefont
  {Grandjean}, \citenamefont {Damilano}, \citenamefont {Dalmasso},
  \citenamefont {Leroux}, \citenamefont {La{\"u}gt},\ and\ \citenamefont
  {Massies}}]{Grandjean1999}%
  \BibitemOpen
  \bibfield  {author} {\bibinfo {author} {\bibfnamefont {N}~\bibnamefont
  {Grandjean}}, \bibinfo {author} {\bibfnamefont {B}~\bibnamefont {Damilano}},
  \bibinfo {author} {\bibfnamefont {S}~\bibnamefont {Dalmasso}}, \bibinfo
  {author} {\bibfnamefont {M}~\bibnamefont {Leroux}}, \bibinfo {author}
  {\bibfnamefont {M}~\bibnamefont {La{\"u}gt}}, \ and\ \bibinfo {author}
  {\bibfnamefont {J}~\bibnamefont {Massies}},\ }\bibfield  {title} {\enquote
  {\bibinfo {title} {{Built-in electric-field effects in wurtzite AlGaN/GaN
  quantum wells}},}\ }\href@noop {} {\bibfield  {journal} {\bibinfo  {journal}
  {J. Appl. Phys.}\ }\textbf {\bibinfo {volume} {86}},\ \bibinfo {pages} {3714}
  (\bibinfo {year} {1999})}\BibitemShut {NoStop}%
\bibitem [{\citenamefont {Kash}\ \emph {et~al.}(1991)\citenamefont {Kash},
  \citenamefont {Zachau}, \citenamefont {Mendez}, \citenamefont {Hong},\ and\
  \citenamefont {Fukuzawa}}]{Kash1991}%
  \BibitemOpen
  \bibfield  {author} {\bibinfo {author} {\bibfnamefont {J~A}\ \bibnamefont
  {Kash}}, \bibinfo {author} {\bibfnamefont {M}~\bibnamefont {Zachau}},
  \bibinfo {author} {\bibfnamefont {E~E}\ \bibnamefont {Mendez}}, \bibinfo
  {author} {\bibfnamefont {J~M}\ \bibnamefont {Hong}}, \ and\ \bibinfo {author}
  {\bibfnamefont {T}~\bibnamefont {Fukuzawa}},\ }\bibfield  {title} {\enquote
  {\bibinfo {title} {{Fermi-Dirac distribution of excitons in coupled quantum
  wells}},}\ }\href@noop {} {\bibfield  {journal} {\bibinfo  {journal} {Phys.
  Rev. Lett.}\ }\textbf {\bibinfo {volume} {66}},\ \bibinfo {pages}
  {2247--2250} (\bibinfo {year} {1991})}\BibitemShut {NoStop}%
\bibitem [{\citenamefont {Schindler}\ and\ \citenamefont
  {Zimmermann}(2008)}]{Zimmermann}%
  \BibitemOpen
  \bibfield  {author} {\bibinfo {author} {\bibfnamefont {Christoph}\
  \bibnamefont {Schindler}}\ and\ \bibinfo {author} {\bibfnamefont {Roland}\
  \bibnamefont {Zimmermann}},\ }\bibfield  {title} {\enquote {\bibinfo {title}
  {Analysis of the exciton-exciton interaction in semiconductor quantum
  wells},}\ }\href@noop {} {\bibfield  {journal} {\bibinfo  {journal} {Phys.
  Rev. B}\ }\textbf {\bibinfo {volume} {78}},\ \bibinfo {pages} {045313}
  (\bibinfo {year} {2008})}\BibitemShut {NoStop}%
\bibitem [{\citenamefont {Laikhtman}\ and\ \citenamefont
  {Rapaport}(2009{\natexlab{b}})}]{Laikhtman}%
  \BibitemOpen
  \bibfield  {author} {\bibinfo {author} {\bibfnamefont {B.}~\bibnamefont
  {Laikhtman}}\ and\ \bibinfo {author} {\bibfnamefont {R.}~\bibnamefont
  {Rapaport}},\ }\bibfield  {title} {\enquote {\bibinfo {title} {{Correlations
  in a two-dimensional Bose gas with long-range interaction}},}\ }\href@noop {}
  {\bibfield  {journal} {\bibinfo  {journal} {Europhysics Letters}\ }\textbf
  {\bibinfo {volume} {87}},\ \bibinfo {pages} {27010} (\bibinfo {year}
  {2009}{\natexlab{b}})}\BibitemShut {NoStop}%
\bibitem [{\citenamefont {Mazuz-Harpaz}\ \emph {et~al.}(2017)\citenamefont
  {Mazuz-Harpaz}, \citenamefont {Cohen},\ and\ \citenamefont
  {Rapaport}}]{MazuzHarpaz2017}%
  \BibitemOpen
  \bibfield  {author} {\bibinfo {author} {\bibfnamefont {Yotam}\ \bibnamefont
  {Mazuz-Harpaz}}, \bibinfo {author} {\bibfnamefont {Kobi}\ \bibnamefont
  {Cohen}}, \ and\ \bibinfo {author} {\bibfnamefont {Ronen}\ \bibnamefont
  {Rapaport}},\ }\bibfield  {title} {\enquote {\bibinfo {title} {{Condensation
  to a strongly correlated dark fluid of two dimensional dipolar excitons}},}\
  }\href@noop {} {\bibfield  {journal} {\bibinfo  {journal} {Superlattices and
  Microstructures}\ }\textbf {\bibinfo {volume} {108}},\ \bibinfo {pages}
  {88--97} (\bibinfo {year} {2017})}\BibitemShut {NoStop}%
\bibitem [{\citenamefont {Zimmermann}(1987)}]{ZimmermannBook}%
  \BibitemOpen
  \bibfield  {author} {\bibinfo {author} {\bibfnamefont {Roland}\ \bibnamefont
  {Zimmermann}},\ }\href@noop {} {\emph {\bibinfo {title} {Many-Particle Theory
  of Highly Excited Semiconductors}}}\ (\bibinfo  {publisher} {BSB Teubner},\
  \bibinfo {year} {1987})\BibitemShut {NoStop}%
\bibitem [{Note1()}]{Note1}%
  \BibitemOpen
  \bibinfo {note} {Note, that so-called magnetoexcitons characterised by linear
  energy shift under magnetic field are routinely observed in GaAs when
  magnetic length becomes smaller than exciton Bohr radius. This regime is
  never achieved in this work}\BibitemShut {NoStop}%
\bibitem [{\citenamefont {Kuznetsova}\ \emph {et~al.}(2017)\citenamefont
  {Kuznetsova}, \citenamefont {Dorow}, \citenamefont {Calman}, \citenamefont
  {Butov}, \citenamefont {Wilkes}, \citenamefont {Muljarov}, \citenamefont
  {Campman},\ and\ \citenamefont {Gossard}}]{Kuznetsova2017}%
  \BibitemOpen
  \bibfield  {author} {\bibinfo {author} {\bibfnamefont {Y~Y}\ \bibnamefont
  {Kuznetsova}}, \bibinfo {author} {\bibfnamefont {C~J}\ \bibnamefont {Dorow}},
  \bibinfo {author} {\bibfnamefont {E~V}\ \bibnamefont {Calman}}, \bibinfo
  {author} {\bibfnamefont {L~V}\ \bibnamefont {Butov}}, \bibinfo {author}
  {\bibfnamefont {J}~\bibnamefont {Wilkes}}, \bibinfo {author} {\bibfnamefont
  {E~A}\ \bibnamefont {Muljarov}}, \bibinfo {author} {\bibfnamefont {K~L}\
  \bibnamefont {Campman}}, \ and\ \bibinfo {author} {\bibfnamefont {A~C}\
  \bibnamefont {Gossard}},\ }\bibfield  {title} {\enquote {\bibinfo {title}
  {{Transport of indirect excitons in high magnetic fields}},}\ }\href@noop {}
  {\bibfield  {journal} {\bibinfo  {journal} {Phys. Rev. B}\ }\textbf {\bibinfo
  {volume} {95}},\ \bibinfo {pages} {125304} (\bibinfo {year}
  {2017})}\BibitemShut {NoStop}%
\bibitem [{\citenamefont {Dorow}\ \emph {et~al.}(2017)\citenamefont {Dorow},
  \citenamefont {Hasling}, \citenamefont {Calman}, \citenamefont {Butov},
  \citenamefont {Wilkes}, \citenamefont {Campman},\ and\ \citenamefont
  {Gossard}}]{Dorow2017}%
  \BibitemOpen
  \bibfield  {author} {\bibinfo {author} {\bibfnamefont {C~J}\ \bibnamefont
  {Dorow}}, \bibinfo {author} {\bibfnamefont {M~W}\ \bibnamefont {Hasling}},
  \bibinfo {author} {\bibfnamefont {E~V}\ \bibnamefont {Calman}}, \bibinfo
  {author} {\bibfnamefont {L~V}\ \bibnamefont {Butov}}, \bibinfo {author}
  {\bibfnamefont {J}~\bibnamefont {Wilkes}}, \bibinfo {author} {\bibfnamefont
  {K~L}\ \bibnamefont {Campman}}, \ and\ \bibinfo {author} {\bibfnamefont
  {A~C}\ \bibnamefont {Gossard}},\ }\bibfield  {title} {\enquote {\bibinfo
  {title} {{Spatially resolved and time-resolved imaging of transport of
  indirect excitons in high magnetic fields}},}\ }\href@noop {} {\bibfield
  {journal} {\bibinfo  {journal} {Phys. Rev. B}\ }\textbf {\bibinfo {volume}
  {95}},\ \bibinfo {pages} {235308} (\bibinfo {year} {2017})}\BibitemShut
  {NoStop}%
\bibitem [{\citenamefont {Grandjean}\ \emph {et~al.}(2000)\citenamefont
  {Grandjean}, \citenamefont {Damilano}, \citenamefont {Massies}, \citenamefont
  {Neu}, \citenamefont {Teissere}, \citenamefont {Grzegory}, \citenamefont
  {Porowski}, \citenamefont {Gallart}, \citenamefont {Lefebvre}, \citenamefont
  {Gil},\ and\ \citenamefont {Albrecht}}]{Grandjean2000}%
  \BibitemOpen
  \bibfield  {author} {\bibinfo {author} {\bibfnamefont {N}~\bibnamefont
  {Grandjean}}, \bibinfo {author} {\bibfnamefont {B}~\bibnamefont {Damilano}},
  \bibinfo {author} {\bibfnamefont {J}~\bibnamefont {Massies}}, \bibinfo
  {author} {\bibfnamefont {G}~\bibnamefont {Neu}}, \bibinfo {author}
  {\bibfnamefont {M}~\bibnamefont {Teissere}}, \bibinfo {author} {\bibfnamefont
  {I}~\bibnamefont {Grzegory}}, \bibinfo {author} {\bibfnamefont
  {S}~\bibnamefont {Porowski}}, \bibinfo {author} {\bibfnamefont
  {M}~\bibnamefont {Gallart}}, \bibinfo {author} {\bibfnamefont
  {P}~\bibnamefont {Lefebvre}}, \bibinfo {author} {\bibfnamefont
  {B}~\bibnamefont {Gil}}, \ and\ \bibinfo {author} {\bibfnamefont
  {M}~\bibnamefont {Albrecht}},\ }\bibfield  {title} {\enquote {\bibinfo
  {title} {{Optical properties of GaN epilayers and GaN/AlGaN quantum wells
  grown by molecular beam epitaxy on GaN(0001) single crystal substrate}},}\
  }\href@noop {} {\bibfield  {journal} {\bibinfo  {journal} {J. Appl. Phys.}\
  }\textbf {\bibinfo {volume} {88}},\ \bibinfo {pages} {183--187} (\bibinfo
  {year} {2000})}\BibitemShut {NoStop}%
\bibitem [{\citenamefont {Lefebvre}\ \emph {et~al.}(1999)\citenamefont
  {Lefebvre}, \citenamefont {All{\`e}gre}, \citenamefont {Gil}, \citenamefont
  {Mathieu}, \citenamefont {Grandjean}, \citenamefont {Leroux}, \citenamefont
  {Massies},\ and\ \citenamefont {Bigenwald}}]{Lefebvre1999}%
  \BibitemOpen
  \bibfield  {author} {\bibinfo {author} {\bibfnamefont {Pierre}\ \bibnamefont
  {Lefebvre}}, \bibinfo {author} {\bibfnamefont {Jacques}\ \bibnamefont
  {All{\`e}gre}}, \bibinfo {author} {\bibfnamefont {Bernard}\ \bibnamefont
  {Gil}}, \bibinfo {author} {\bibfnamefont {Henry}\ \bibnamefont {Mathieu}},
  \bibinfo {author} {\bibfnamefont {Nicolas}\ \bibnamefont {Grandjean}},
  \bibinfo {author} {\bibfnamefont {Mathieu}\ \bibnamefont {Leroux}}, \bibinfo
  {author} {\bibfnamefont {Jean}\ \bibnamefont {Massies}}, \ and\ \bibinfo
  {author} {\bibfnamefont {Pierre}\ \bibnamefont {Bigenwald}},\ }\bibfield
  {title} {\enquote {\bibinfo {title} {{Time-resolved photoluminescence as a
  probe of internal electric fields in GaN-(GaAl)N quantum wells}},}\
  }\href@noop {} {\bibfield  {journal} {\bibinfo  {journal} {Phys. Rev. B}\
  }\textbf {\bibinfo {volume} {59}},\ \bibinfo {pages} {15363--15367} (\bibinfo
  {year} {1999})}\BibitemShut {NoStop}%
\bibitem [{\citenamefont {Fedichkin}\ \emph {et~al.}(2015)\citenamefont
  {Fedichkin}, \citenamefont {Andreakou}, \citenamefont {Jouault},
  \citenamefont {Vladimirova}, \citenamefont {Guillet}, \citenamefont
  {Brimont}, \citenamefont {Valvin}, \citenamefont {Bretagnon}, \citenamefont
  {Dussaigne}, \citenamefont {Grandjean},\ and\ \citenamefont
  {Lefebvre}}]{Fedichkin2016}%
  \BibitemOpen
  \bibfield  {author} {\bibinfo {author} {\bibfnamefont {F.}~\bibnamefont
  {Fedichkin}}, \bibinfo {author} {\bibfnamefont {P.}~\bibnamefont
  {Andreakou}}, \bibinfo {author} {\bibfnamefont {B.}~\bibnamefont {Jouault}},
  \bibinfo {author} {\bibfnamefont {M.}~\bibnamefont {Vladimirova}}, \bibinfo
  {author} {\bibfnamefont {T.}~\bibnamefont {Guillet}}, \bibinfo {author}
  {\bibfnamefont {C.}~\bibnamefont {Brimont}}, \bibinfo {author} {\bibfnamefont
  {P.}~\bibnamefont {Valvin}}, \bibinfo {author} {\bibfnamefont
  {T.}~\bibnamefont {Bretagnon}}, \bibinfo {author} {\bibfnamefont
  {A.}~\bibnamefont {Dussaigne}}, \bibinfo {author} {\bibfnamefont
  {N.}~\bibnamefont {Grandjean}}, \ and\ \bibinfo {author} {\bibfnamefont
  {P.}~\bibnamefont {Lefebvre}},\ }\bibfield  {title} {\enquote {\bibinfo
  {title} {{Transport of dipolar excitons in (Al,Ga)N/GaN quantum wells}},}\
  }\href@noop {} {\bibfield  {journal} {\bibinfo  {journal} {Phys. Rev. B}\
  }\textbf {\bibinfo {volume} {91}},\ \bibinfo {pages} {205424} (\bibinfo
  {year} {2015})}\BibitemShut {NoStop}%
\bibitem [{\citenamefont {Rosales}\ \emph {et~al.}(2013)\citenamefont
  {Rosales}, \citenamefont {Bretagnon}, \citenamefont {Gil}, \citenamefont
  {Kahouli}, \citenamefont {Brault}, \citenamefont {Damilano}, \citenamefont
  {Massies}, \citenamefont {Durnev},\ and\ \citenamefont
  {Kavokin}}]{Rosales2013}%
  \BibitemOpen
  \bibfield  {author} {\bibinfo {author} {\bibfnamefont {D}~\bibnamefont
  {Rosales}}, \bibinfo {author} {\bibfnamefont {T}~\bibnamefont {Bretagnon}},
  \bibinfo {author} {\bibfnamefont {B}~\bibnamefont {Gil}}, \bibinfo {author}
  {\bibfnamefont {A}~\bibnamefont {Kahouli}}, \bibinfo {author} {\bibfnamefont
  {J}~\bibnamefont {Brault}}, \bibinfo {author} {\bibfnamefont {B}~\bibnamefont
  {Damilano}}, \bibinfo {author} {\bibfnamefont {J}~\bibnamefont {Massies}},
  \bibinfo {author} {\bibfnamefont {M~V}\ \bibnamefont {Durnev}}, \ and\
  \bibinfo {author} {\bibfnamefont {A~V}\ \bibnamefont {Kavokin}},\ }\bibfield
  {title} {\enquote {\bibinfo {title} {{Excitons in nitride heterostructures:
  From zero- to one-dimensional behavior}},}\ }\href@noop {} {\bibfield
  {journal} {\bibinfo  {journal} {Phys. Rev. B}\ }\textbf {\bibinfo {volume}
  {88}},\ \bibinfo {pages} {125437} (\bibinfo {year} {2013})}\BibitemShut
  {NoStop}%
\bibitem [{\citenamefont {Rossbach}\ \emph {et~al.}(2014)\citenamefont
  {Rossbach}, \citenamefont {Levrat}, \citenamefont {Jacopin}, \citenamefont
  {Shahmohammadi}, \citenamefont {Carlin}, \citenamefont {Gani{\`e}re},
  \citenamefont {Butt{\'e}}, \citenamefont {Deveaud},\ and\ \citenamefont
  {Grandjean}}]{Rossbach2014}%
  \BibitemOpen
  \bibfield  {author} {\bibinfo {author} {\bibfnamefont {G}~\bibnamefont
  {Rossbach}}, \bibinfo {author} {\bibfnamefont {J}~\bibnamefont {Levrat}},
  \bibinfo {author} {\bibfnamefont {G}~\bibnamefont {Jacopin}}, \bibinfo
  {author} {\bibfnamefont {M}~\bibnamefont {Shahmohammadi}}, \bibinfo {author}
  {\bibfnamefont {J~F}\ \bibnamefont {Carlin}}, \bibinfo {author}
  {\bibfnamefont {J~D}\ \bibnamefont {Gani{\`e}re}}, \bibinfo {author}
  {\bibfnamefont {R}~\bibnamefont {Butt{\'e}}}, \bibinfo {author}
  {\bibfnamefont {B}~\bibnamefont {Deveaud}}, \ and\ \bibinfo {author}
  {\bibfnamefont {N}~\bibnamefont {Grandjean}},\ }\bibfield  {title} {\enquote
  {\bibinfo {title} {{High-temperature Mott transition in wide-band-gap
  semiconductor quantum wells}},}\ }\href@noop {} {\bibfield  {journal}
  {\bibinfo  {journal} {Phys. Rev. B}\ }\textbf {\bibinfo {volume} {90}},\
  \bibinfo {pages} {201308} (\bibinfo {year} {2014})}\BibitemShut {NoStop}%
\bibitem [{\citenamefont {Fedichkin}\ \emph {et~al.}(2016)\citenamefont
  {Fedichkin}, \citenamefont {Guillet}, \citenamefont {Valvin}, \citenamefont
  {Jouault}, \citenamefont {Brimont}, \citenamefont {Bretagnon}, \citenamefont
  {Lahourcade}, \citenamefont {Grandjean}, \citenamefont {Lefebvre},\ and\
  \citenamefont {Vladimirova}}]{Fedichkin2015}%
  \BibitemOpen
  \bibfield  {author} {\bibinfo {author} {\bibfnamefont {F}~\bibnamefont
  {Fedichkin}}, \bibinfo {author} {\bibfnamefont {T}~\bibnamefont {Guillet}},
  \bibinfo {author} {\bibfnamefont {P}~\bibnamefont {Valvin}}, \bibinfo
  {author} {\bibfnamefont {B}~\bibnamefont {Jouault}}, \bibinfo {author}
  {\bibfnamefont {C}~\bibnamefont {Brimont}}, \bibinfo {author} {\bibfnamefont
  {T}~\bibnamefont {Bretagnon}}, \bibinfo {author} {\bibfnamefont
  {L}~\bibnamefont {Lahourcade}}, \bibinfo {author} {\bibfnamefont
  {N}~\bibnamefont {Grandjean}}, \bibinfo {author} {\bibfnamefont
  {P}~\bibnamefont {Lefebvre}}, \ and\ \bibinfo {author} {\bibfnamefont
  {M}~\bibnamefont {Vladimirova}},\ }\bibfield  {title} {\enquote {\bibinfo
  {title} {{Room-Temperature Transport of Indirect Excitons in (Al,Ga)N/GaN
  Quantum Wells}},}\ }\href@noop {} {\bibfield  {journal} {\bibinfo  {journal}
  {Phys. Rev. Applied}\ }\textbf {\bibinfo {volume} {6}},\ \bibinfo {pages}
  {014011} (\bibinfo {year} {2016})}\BibitemShut {NoStop}%
\bibitem [{\citenamefont {Lefebvre}\ \emph {et~al.}(2004)\citenamefont
  {Lefebvre}, \citenamefont {Kalliakos}, \citenamefont {Bretagnon},
  \citenamefont {Valvin}, \citenamefont {Taliercio}, \citenamefont {Gil},
  \citenamefont {Grandjean},\ and\ \citenamefont {Massies}}]{Lefebvre2004}%
  \BibitemOpen
  \bibfield  {author} {\bibinfo {author} {\bibfnamefont {P}~\bibnamefont
  {Lefebvre}}, \bibinfo {author} {\bibfnamefont {S}~\bibnamefont {Kalliakos}},
  \bibinfo {author} {\bibfnamefont {T}~\bibnamefont {Bretagnon}}, \bibinfo
  {author} {\bibfnamefont {P}~\bibnamefont {Valvin}}, \bibinfo {author}
  {\bibfnamefont {T}~\bibnamefont {Taliercio}}, \bibinfo {author}
  {\bibfnamefont {B}~\bibnamefont {Gil}}, \bibinfo {author} {\bibfnamefont
  {N}~\bibnamefont {Grandjean}}, \ and\ \bibinfo {author} {\bibfnamefont
  {J}~\bibnamefont {Massies}},\ }\bibfield  {title} {\enquote {\bibinfo {title}
  {{Observation and modeling of the time-dependent descreening of internal
  electric field in a wurtzite GaN/Al$_{0.15}$Ga$_{0.85}$N quantum well after
  high photoexcitation}},}\ }\href@noop {} {\bibfield  {journal} {\bibinfo
  {journal} {Phys. Rev. B}\ }\textbf {\bibinfo {volume} {69}},\ \bibinfo
  {pages} {035307} (\bibinfo {year} {2004})}\BibitemShut {NoStop}%
\bibitem [{\citenamefont {Remeika}\ \emph {et~al.}(2009)\citenamefont
  {Remeika}, \citenamefont {Graves}, \citenamefont {Hammack}, \citenamefont
  {Meyertholen}, \citenamefont {Fogler}, \citenamefont {Butov}, \citenamefont
  {Hanson},\ and\ \citenamefont {Gossard}}]{Remeika2009}%
  \BibitemOpen
  \bibfield  {author} {\bibinfo {author} {\bibfnamefont {M}~\bibnamefont
  {Remeika}}, \bibinfo {author} {\bibfnamefont {J~C}\ \bibnamefont {Graves}},
  \bibinfo {author} {\bibfnamefont {A~T}\ \bibnamefont {Hammack}}, \bibinfo
  {author} {\bibfnamefont {A~D}\ \bibnamefont {Meyertholen}}, \bibinfo {author}
  {\bibfnamefont {M~M}\ \bibnamefont {Fogler}}, \bibinfo {author}
  {\bibfnamefont {L~V}\ \bibnamefont {Butov}}, \bibinfo {author} {\bibfnamefont
  {M}~\bibnamefont {Hanson}}, \ and\ \bibinfo {author} {\bibfnamefont {A~C}\
  \bibnamefont {Gossard}},\ }\bibfield  {title} {\enquote {\bibinfo {title}
  {{Localization-Delocalization Transition of Indirect Excitons in Lateral
  Electrostatic Lattices}},}\ }\href@noop {} {\bibfield  {journal} {\bibinfo
  {journal} {Phys. Rev. Lett.}\ }\textbf {\bibinfo {volume} {102}},\ \bibinfo
  {pages} {186803--4} (\bibinfo {year} {2009})}\BibitemShut {NoStop}%
\bibitem [{\citenamefont {Leonard}\ \emph {et~al.}(2012)\citenamefont
  {Leonard}, \citenamefont {Remeika}, \citenamefont {Chu}, \citenamefont
  {Kuznetsova}, \citenamefont {High}, \citenamefont {Butov}, \citenamefont
  {Wilkes}, \citenamefont {Hanson},\ and\ \citenamefont
  {Gossard}}]{Leonard2012}%
  \BibitemOpen
  \bibfield  {author} {\bibinfo {author} {\bibfnamefont {J~R}\ \bibnamefont
  {Leonard}}, \bibinfo {author} {\bibfnamefont {M}~\bibnamefont {Remeika}},
  \bibinfo {author} {\bibfnamefont {M~K}\ \bibnamefont {Chu}}, \bibinfo
  {author} {\bibfnamefont {Y~Y}\ \bibnamefont {Kuznetsova}}, \bibinfo {author}
  {\bibfnamefont {A~A}\ \bibnamefont {High}}, \bibinfo {author} {\bibfnamefont
  {L~V}\ \bibnamefont {Butov}}, \bibinfo {author} {\bibfnamefont
  {J}~\bibnamefont {Wilkes}}, \bibinfo {author} {\bibfnamefont {M}~\bibnamefont
  {Hanson}}, \ and\ \bibinfo {author} {\bibfnamefont {A~C}\ \bibnamefont
  {Gossard}},\ }\bibfield  {title} {\enquote {\bibinfo {title} {{Transport of
  indirect excitons in a potential energy gradient}},}\ }\href@noop {}
  {\bibfield  {journal} {\bibinfo  {journal} {Appl. Phys. Lett.}\ }\textbf
  {\bibinfo {volume} {100}},\ \bibinfo {pages} {231106--231105} (\bibinfo
  {year} {2012})}\BibitemShut {NoStop}%
\bibitem [{\citenamefont {Rapaport}\ \emph {et~al.}(2006)\citenamefont
  {Rapaport}, \citenamefont {Chen},\ and\ \citenamefont
  {Simon}}]{Rapaport2006}%
  \BibitemOpen
  \bibfield  {author} {\bibinfo {author} {\bibfnamefont {Ronen}\ \bibnamefont
  {Rapaport}}, \bibinfo {author} {\bibfnamefont {Gang}\ \bibnamefont {Chen}}, \
  and\ \bibinfo {author} {\bibfnamefont {Steven~H.}\ \bibnamefont {Simon}},\
  }\bibfield  {title} {\enquote {\bibinfo {title} {Nonlinear dynamics of a
  dense two-dimensional dipolar exciton gas},}\ }\href {\doibase
  10.1103/PhysRevB.73.033319} {\bibfield  {journal} {\bibinfo  {journal} {Phys.
  Rev. B}\ }\textbf {\bibinfo {volume} {73}},\ \bibinfo {pages} {033319}
  (\bibinfo {year} {2006})}\BibitemShut {NoStop}%
\bibitem [{\citenamefont {Ivanov}(2002)}]{Ivanov2002}%
  \BibitemOpen
  \bibfield  {author} {\bibinfo {author} {\bibfnamefont {A~L}\ \bibnamefont
  {Ivanov}},\ }\bibfield  {title} {\enquote {\bibinfo {title} {{Quantum
  diffusion of dipole-oriented indirect excitons in coupled quantum wells}},}\
  }\href@noop {} {\bibfield  {journal} {\bibinfo  {journal} {EPL}\ }\textbf
  {\bibinfo {volume} {59}},\ \bibinfo {pages} {586--591} (\bibinfo {year}
  {2002})}\BibitemShut {NoStop}%
\bibitem [{\citenamefont {Lozovik}\ and\ \citenamefont
  {Ruvinskii}(1997)}]{Lozovik1997}%
  \BibitemOpen
  \bibfield  {author} {\bibinfo {author} {\bibfnamefont {Yu~E}\ \bibnamefont
  {Lozovik}}\ and\ \bibinfo {author} {\bibfnamefont {A~M}\ \bibnamefont
  {Ruvinskii}},\ }\bibfield  {title} {\enquote {\bibinfo {title}
  {{Magnetoexciton absorption in coupled quantum wells}},}\ }\href@noop {}
  {\bibfield  {journal} {\bibinfo  {journal} {J. Exp. Theor. Phys.}\ }\textbf
  {\bibinfo {volume} {85}},\ \bibinfo {pages} {979--988} (\bibinfo {year}
  {1997})}\BibitemShut {NoStop}%
\bibitem [{\citenamefont {Butov}\ \emph {et~al.}(2001)\citenamefont {Butov},
  \citenamefont {Lai}, \citenamefont {Chemla}, \citenamefont {Lozovik},
  \citenamefont {Campman},\ and\ \citenamefont {Gossard}}]{Butov2001}%
  \BibitemOpen
  \bibfield  {author} {\bibinfo {author} {\bibfnamefont {L~V}\ \bibnamefont
  {Butov}}, \bibinfo {author} {\bibfnamefont {C~W}\ \bibnamefont {Lai}},
  \bibinfo {author} {\bibfnamefont {D~S}\ \bibnamefont {Chemla}}, \bibinfo
  {author} {\bibfnamefont {Yu~E}\ \bibnamefont {Lozovik}}, \bibinfo {author}
  {\bibfnamefont {K~L}\ \bibnamefont {Campman}}, \ and\ \bibinfo {author}
  {\bibfnamefont {A~C}\ \bibnamefont {Gossard}},\ }\bibfield  {title} {\enquote
  {\bibinfo {title} {{Observation of Magnetically Induced Effective-Mass
  Enhancement of Quasi-2D Excitons}},}\ }\href@noop {} {\bibfield  {journal}
  {\bibinfo  {journal} {Phys. Rev. Lett.}\ }\textbf {\bibinfo {volume} {87}},\
  \bibinfo {pages} {216804} (\bibinfo {year} {2001})}\BibitemShut {NoStop}%
\bibitem [{\citenamefont {Lozovik}\ \emph {et~al.}(2002)\citenamefont
  {Lozovik}, \citenamefont {Ovchinnikov}, \citenamefont {Volkov}, \citenamefont
  {Butov},\ and\ \citenamefont {Chemla}}]{Lozovik2002}%
  \BibitemOpen
  \bibfield  {author} {\bibinfo {author} {\bibfnamefont {Yu~E}\ \bibnamefont
  {Lozovik}}, \bibinfo {author} {\bibfnamefont {I~V}\ \bibnamefont
  {Ovchinnikov}}, \bibinfo {author} {\bibfnamefont {S~Yu}\ \bibnamefont
  {Volkov}}, \bibinfo {author} {\bibfnamefont {L~V}\ \bibnamefont {Butov}}, \
  and\ \bibinfo {author} {\bibfnamefont {D~S}\ \bibnamefont {Chemla}},\
  }\bibfield  {title} {\enquote {\bibinfo {title} {{Quasi-two-dimensional
  excitons in finite magnetic fields}},}\ }\href@noop {} {\bibfield  {journal}
  {\bibinfo  {journal} {Phys. Rev. B}\ }\textbf {\bibinfo {volume} {65}},\
  \bibinfo {pages} {235304} (\bibinfo {year} {2002})}\BibitemShut {NoStop}%
\bibitem [{\citenamefont {Edelstein}\ \emph {et~al.}(1989)\citenamefont
  {Edelstein}, \citenamefont {Spector},\ and\ \citenamefont
  {Marasas}}]{Edelstein}%
  \BibitemOpen
  \bibfield  {author} {\bibinfo {author} {\bibfnamefont {Warren}\ \bibnamefont
  {Edelstein}}, \bibinfo {author} {\bibfnamefont {Harold~N}\ \bibnamefont
  {Spector}}, \ and\ \bibinfo {author} {\bibfnamefont {Richard}\ \bibnamefont
  {Marasas}},\ }\bibfield  {title} {\enquote {\bibinfo {title}
  {{Two-dimensional excitons in magnetic fields}},}\ }\href@noop {} {\bibfield
  {journal} {\bibinfo  {journal} {Phys. Rev. B}\ }\textbf {\bibinfo {volume}
  {39}},\ \bibinfo {pages} {7697--7704} (\bibinfo {year} {1989})}\BibitemShut
  {NoStop}%
\bibitem [{\citenamefont {St{\'e}pnicki}\ \emph {et~al.}(2015)\citenamefont
  {St{\'e}pnicki}, \citenamefont {Pi{\'e}tka}, \citenamefont {Morier-Genoud},
  \citenamefont {Deveaud},\ and\ \citenamefont {Matuszewski}}]{Stepnicki2015}%
  \BibitemOpen
  \bibfield  {author} {\bibinfo {author} {\bibfnamefont {Piotr}\ \bibnamefont
  {St{\'e}pnicki}}, \bibinfo {author} {\bibfnamefont {Barbara}\ \bibnamefont
  {Pi{\'e}tka}}, \bibinfo {author} {\bibfnamefont {Fran{\c c}ois}\ \bibnamefont
  {Morier-Genoud}}, \bibinfo {author} {\bibfnamefont {Benoit}\ \bibnamefont
  {Deveaud}}, \ and\ \bibinfo {author} {\bibfnamefont {Micha{\l}}\ \bibnamefont
  {Matuszewski}},\ }\bibfield  {title} {\enquote {\bibinfo {title} {{Analytical
  method for determining quantum well exciton properties in a magnetic
  field}},}\ }\href@noop {} {\bibfield  {journal} {\bibinfo  {journal} {Phys.
  Rev. B}\ }\textbf {\bibinfo {volume} {91}},\ \bibinfo {pages} {195302}
  (\bibinfo {year} {2015})}\BibitemShut {NoStop}%
\bibitem [{\citenamefont {Arnardottir}\ \emph {et~al.}(2012)\citenamefont
  {Arnardottir}, \citenamefont {Kyriienko},\ and\ \citenamefont
  {Shelykh}}]{Arnardottir}%
  \BibitemOpen
  \bibfield  {author} {\bibinfo {author} {\bibfnamefont {K~B}\ \bibnamefont
  {Arnardottir}}, \bibinfo {author} {\bibfnamefont {O}~\bibnamefont
  {Kyriienko}}, \ and\ \bibinfo {author} {\bibfnamefont {I~A}\ \bibnamefont
  {Shelykh}},\ }\bibfield  {title} {\enquote {\bibinfo {title} {{Hall effect
  for indirect excitons in an inhomogeneous magnetic field}},}\ }\href@noop {}
  {\bibfield  {journal} {\bibinfo  {journal} {Phys. Rev. B}\ }\textbf {\bibinfo
  {volume} {86}},\ \bibinfo {pages} {245311} (\bibinfo {year}
  {2012})}\BibitemShut {NoStop}%
\bibitem [{\citenamefont {Wilkes}\ and\ \citenamefont
  {Muljarov}(2017)}]{Wilkes2017}%
  \BibitemOpen
  \bibfield  {author} {\bibinfo {author} {\bibfnamefont {J}~\bibnamefont
  {Wilkes}}\ and\ \bibinfo {author} {\bibfnamefont {E~A}\ \bibnamefont
  {Muljarov}},\ }\bibfield  {title} {\enquote {\bibinfo {title} {{Excitons and
  polaritons in planar heterostructures in external electric and magnetic
  fields: A multi-sub-level approach}},}\ }\href@noop {} {\bibfield  {journal}
  {\bibinfo  {journal} {Superlattices and Microstructures}\ } (\bibinfo {year}
  {2017})}\BibitemShut {NoStop}%
\bibitem [{\citenamefont {Wilkes}\ and\ \citenamefont
  {Muljarov}(2016)}]{Wilkes2016}%
  \BibitemOpen
  \bibfield  {author} {\bibinfo {author} {\bibfnamefont {J}~\bibnamefont
  {Wilkes}}\ and\ \bibinfo {author} {\bibfnamefont {E~A}\ \bibnamefont
  {Muljarov}},\ }\bibfield  {title} {\enquote {\bibinfo {title} {{Exciton
  effective mass enhancement in coupled quantum wells in electric and magnetic
  fields}},}\ }\href@noop {} {\bibfield  {journal} {\bibinfo  {journal} {New J.
  Phys.}\ }\textbf {\bibinfo {volume} {18}},\ \bibinfo {pages} {023032}
  (\bibinfo {year} {2016})}\BibitemShut {NoStop}%
\bibitem [{\citenamefont {Arseev}\ and\ \citenamefont
  {Dzyubenko}(1998)}]{Arseev1998}%
  \BibitemOpen
  \bibfield  {author} {\bibinfo {author} {\bibfnamefont {P~I}\ \bibnamefont
  {Arseev}}\ and\ \bibinfo {author} {\bibfnamefont {A~B}\ \bibnamefont
  {Dzyubenko}},\ }\bibfield  {title} {\enquote {\bibinfo {title} {{Exciton
  magnetotransport in two-dimensional systems: Weak-localization effects}},}\
  }\href@noop {} {\bibfield  {journal} {\bibinfo  {journal} {J. Exp. Theor.
  Phys.}\ }\textbf {\bibinfo {volume} {87}},\ \bibinfo {pages} {200--209}
  (\bibinfo {year} {1998})}\BibitemShut {NoStop}%
\bibitem [{\citenamefont {Bugajski}\ \emph {et~al.}(1986)\citenamefont
  {Bugajski}, \citenamefont {Kuszko},\ and\ \citenamefont
  {Regi{\'{n}}ski}}]{Bugajski}%
  \BibitemOpen
  \bibfield  {author} {\bibinfo {author} {\bibfnamefont {M}~\bibnamefont
  {Bugajski}}, \bibinfo {author} {\bibfnamefont {W}~\bibnamefont {Kuszko}}, \
  and\ \bibinfo {author} {\bibfnamefont {K}~\bibnamefont {Regi{\'{n}}ski}},\
  }\bibfield  {title} {\enquote {\bibinfo {title} {{Diamagnetic shift of
  exciton energy levels in GaAs-GaAlAs quantum wells}},}\ }\href@noop {}
  {\bibfield  {journal} {\bibinfo  {journal} {Solid State Communications}\
  }\textbf {\bibinfo {volume} {60}},\ \bibinfo {pages} {669} (\bibinfo {year}
  {1986})}\BibitemShut {NoStop}%
\bibitem [{\citenamefont {Bigenwald}\ \emph {et~al.}(1999)\citenamefont
  {Bigenwald}, \citenamefont {Lefebvre}, \citenamefont {Bretagnon},\ and\
  \citenamefont {Gil}}]{Bigenwald1999}%
  \BibitemOpen
  \bibfield  {author} {\bibinfo {author} {\bibfnamefont {P}~\bibnamefont
  {Bigenwald}}, \bibinfo {author} {\bibfnamefont {P}~\bibnamefont {Lefebvre}},
  \bibinfo {author} {\bibfnamefont {T}~\bibnamefont {Bretagnon}}, \ and\
  \bibinfo {author} {\bibfnamefont {B}~\bibnamefont {Gil}},\ }\bibfield
  {title} {\enquote {\bibinfo {title} {{Confined Excitons in
  GaN{\textendash}AlGaN Quantum Wells}},}\ }\href@noop {} {\bibfield  {journal}
  {\bibinfo  {journal} {Phys. Status solidi (b)}\ }\textbf {\bibinfo {volume}
  {216}},\ \bibinfo {pages} {371--374} (\bibinfo {year} {1999})}\BibitemShut
  {NoStop}%
\bibitem [{\citenamefont {Bigenwald}\ \emph {et~al.}()\citenamefont
  {Bigenwald}, \citenamefont {Kavokin}, \citenamefont {Gil},\ and\
  \citenamefont {Lefebvre}}]{Bigenwald2001}%
  \BibitemOpen
  \bibfield  {author} {\bibinfo {author} {\bibfnamefont {Pierre}\ \bibnamefont
  {Bigenwald}}, \bibinfo {author} {\bibfnamefont {Alexey}\ \bibnamefont
  {Kavokin}}, \bibinfo {author} {\bibfnamefont {Bernard}\ \bibnamefont {Gil}},
  \ and\ \bibinfo {author} {\bibfnamefont {Pierre}\ \bibnamefont {Lefebvre}},\
  }\bibfield  {title} {\enquote {\bibinfo {title} {{Exclusion principle and
  screening of excitons in GaN/Al$_x$Ga$_{1{\textminus}x}$}n quantum wells},}\
  }\href@noop {} {\ }\BibitemShut {NoStop}%
\bibitem [{\citenamefont {Hangleiter}\ \emph {et~al.}(2015)\citenamefont
  {Hangleiter}, \citenamefont {Jin}, \citenamefont {Gerhard}, \citenamefont
  {Kalincev}, \citenamefont {Langer}, \citenamefont {Bremers}, \citenamefont
  {Rossow}, \citenamefont {Koch}, \citenamefont {Bonn},\ and\ \citenamefont
  {Turchinovich}}]{Hangleiter2015}%
  \BibitemOpen
  \bibfield  {author} {\bibinfo {author} {\bibfnamefont {Andreas}\ \bibnamefont
  {Hangleiter}}, \bibinfo {author} {\bibfnamefont {Zuanming}\ \bibnamefont
  {Jin}}, \bibinfo {author} {\bibfnamefont {Marina}\ \bibnamefont {Gerhard}},
  \bibinfo {author} {\bibfnamefont {Dimitry}\ \bibnamefont {Kalincev}},
  \bibinfo {author} {\bibfnamefont {Torsten}\ \bibnamefont {Langer}}, \bibinfo
  {author} {\bibfnamefont {Heiko}\ \bibnamefont {Bremers}}, \bibinfo {author}
  {\bibfnamefont {Uwe}\ \bibnamefont {Rossow}}, \bibinfo {author}
  {\bibfnamefont {Martin}\ \bibnamefont {Koch}}, \bibinfo {author}
  {\bibfnamefont {Mischa}\ \bibnamefont {Bonn}}, \ and\ \bibinfo {author}
  {\bibfnamefont {Dmitry}\ \bibnamefont {Turchinovich}},\ }\bibfield  {title}
  {\enquote {\bibinfo {title} {{Efficient formation of excitons in a dense
  electron-hole plasma at room temperature}},}\ }\href@noop {} {\bibfield
  {journal} {\bibinfo  {journal} {Phys. Rev. B}\ }\textbf {\bibinfo {volume}
  {92}},\ \bibinfo {pages} {241305} (\bibinfo {year} {2015})}\BibitemShut
  {NoStop}%
\bibitem [{\citenamefont {Bigenwald}\ \emph {et~al.}(2000)\citenamefont
  {Bigenwald}, \citenamefont {Kavokin}, \citenamefont {Gil},\ and\
  \citenamefont {Lefebvre}}]{Bigenwald2000}%
  \BibitemOpen
  \bibfield  {author} {\bibinfo {author} {\bibfnamefont {Pierre}\ \bibnamefont
  {Bigenwald}}, \bibinfo {author} {\bibfnamefont {Alexey}\ \bibnamefont
  {Kavokin}}, \bibinfo {author} {\bibfnamefont {Bernard}\ \bibnamefont {Gil}},
  \ and\ \bibinfo {author} {\bibfnamefont {Pierre}\ \bibnamefont {Lefebvre}},\
  }\bibfield  {title} {\enquote {\bibinfo {title} {{Electron-hole plasma effect
  on excitons in GaN/Al$_x$Ga$_{1{\textminus}x}$ quantum wells}},}\ }\href@noop
  {} {\bibfield  {journal} {\bibinfo  {journal} {Phys. Rev. B}\ }\textbf
  {\bibinfo {volume} {61}},\ \bibinfo {pages} {15621--15624} (\bibinfo {year}
  {2000})}\BibitemShut {NoStop}%
\bibitem [{\citenamefont {Liu}\ \emph {et~al.}(2016)\citenamefont {Liu},
  \citenamefont {Butt{\'e}}, \citenamefont {Dussaigne}, \citenamefont
  {Grandjean}, \citenamefont {Deveaud},\ and\ \citenamefont
  {Jacopin}}]{Liu2016}%
  \BibitemOpen
  \bibfield  {author} {\bibinfo {author} {\bibfnamefont {W}~\bibnamefont
  {Liu}}, \bibinfo {author} {\bibfnamefont {R}~\bibnamefont {Butt{\'e}}},
  \bibinfo {author} {\bibfnamefont {A}~\bibnamefont {Dussaigne}}, \bibinfo
  {author} {\bibfnamefont {N}~\bibnamefont {Grandjean}}, \bibinfo {author}
  {\bibfnamefont {B}~\bibnamefont {Deveaud}}, \ and\ \bibinfo {author}
  {\bibfnamefont {G}~\bibnamefont {Jacopin}},\ }\bibfield  {title} {\enquote
  {\bibinfo {title} {{Carrier-density-dependent recombination dynamics of
  excitons and electron-hole plasma in m-plane InGaN/GaN quantum wells}},}\
  }\href@noop {} {\bibfield  {journal} {\bibinfo  {journal} {Phys. Rev. B}\
  }\textbf {\bibinfo {volume} {94}},\ \bibinfo {pages} {195411} (\bibinfo
  {year} {2016})}\BibitemShut {NoStop}%
\bibitem [{\citenamefont {Vurgaftman}\ \emph {et~al.}(2001)\citenamefont
  {Vurgaftman}, \citenamefont {Meyer},\ and\ \citenamefont
  {Ram-Mohan}}]{Vurgaftman2001}%
  \BibitemOpen
  \bibfield  {author} {\bibinfo {author} {\bibfnamefont {I}~\bibnamefont
  {Vurgaftman}}, \bibinfo {author} {\bibfnamefont {J~R}\ \bibnamefont {Meyer}},
  \ and\ \bibinfo {author} {\bibfnamefont {L~R}\ \bibnamefont {Ram-Mohan}},\
  }\bibfield  {title} {\enquote {\bibinfo {title} {{Band parameters for
  III{\textendash}V compound semiconductors and their alloys}},}\ }\href@noop
  {} {\bibfield  {journal} {\bibinfo  {journal} {J. Appl. Phys.}\ }\textbf
  {\bibinfo {volume} {89}},\ \bibinfo {pages} {5815--5875} (\bibinfo {year}
  {2001})}\BibitemShut {NoStop}%
\bibitem [{\citenamefont {Vurgaftman}\ and\ \citenamefont
  {Meyer}(2003)}]{Vurgaftman2003}%
  \BibitemOpen
  \bibfield  {author} {\bibinfo {author} {\bibfnamefont {I}~\bibnamefont
  {Vurgaftman}}\ and\ \bibinfo {author} {\bibfnamefont {J~R}\ \bibnamefont
  {Meyer}},\ }\bibfield  {title} {\enquote {\bibinfo {title} {{Band parameters
  for nitrogen-containing semiconductors}},}\ }\href@noop {} {\bibfield
  {journal} {\bibinfo  {journal} {J. Appl. Phys.}\ }\textbf {\bibinfo {volume}
  {94}},\ \bibinfo {pages} {3675--3696} (\bibinfo {year} {2003})}\BibitemShut
  {NoStop}%
\bibitem [{\citenamefont {Kabi}\ \emph {et~al.}(2009)\citenamefont {Kabi},
  \citenamefont {Biswas},\ and\ \citenamefont {Panda}}]{Kabi}%
  \BibitemOpen
  \bibfield  {author} {\bibinfo {author} {\bibfnamefont {S}~\bibnamefont
  {Kabi}}, \bibinfo {author} {\bibfnamefont {D}~\bibnamefont {Biswas}}, \ and\
  \bibinfo {author} {\bibfnamefont {S}~\bibnamefont {Panda}},\ }\bibfield
  {title} {\enquote {\bibinfo {title} {Calculations for the band lineup and
  band offsets of algan/gan qws and effects of electric field on the
  photoluminescence},}\ }in\ \href@noop {} {\emph {\bibinfo {booktitle} {4th
  International Conference on Computers and Devices for Communication
  (CODEC)}}}\ (\bibinfo {year} {2009})\BibitemShut {NoStop}%
\bibitem [{\citenamefont {Rickert}\ \emph {et~al.}(2002)\citenamefont
  {Rickert}, \citenamefont {Ellis}, \citenamefont {Kim}, \citenamefont {Lee},
  \citenamefont {Himpsel}, \citenamefont {Dwikusuma},\ and\ \citenamefont
  {Kuech}}]{Rickert2002}%
  \BibitemOpen
  \bibfield  {author} {\bibinfo {author} {\bibfnamefont {K~A}\ \bibnamefont
  {Rickert}}, \bibinfo {author} {\bibfnamefont {A~B}\ \bibnamefont {Ellis}},
  \bibinfo {author} {\bibfnamefont {Jong~Kyu}\ \bibnamefont {Kim}}, \bibinfo
  {author} {\bibfnamefont {Jong-Lam}\ \bibnamefont {Lee}}, \bibinfo {author}
  {\bibfnamefont {F~J}\ \bibnamefont {Himpsel}}, \bibinfo {author}
  {\bibfnamefont {F}~\bibnamefont {Dwikusuma}}, \ and\ \bibinfo {author}
  {\bibfnamefont {T~F}\ \bibnamefont {Kuech}},\ }\bibfield  {title} {\enquote
  {\bibinfo {title} {{X-ray photoemission determination of the Schottky barrier
  height of metal contacts to n{\textendash}GaN and p{\textendash}GaN}},}\
  }\href@noop {} {\bibfield  {journal} {\bibinfo  {journal} {J. Appl. Phys.}\
  }\textbf {\bibinfo {volume} {92}},\ \bibinfo {pages} {6671--6678} (\bibinfo
  {year} {2002})}\BibitemShut {NoStop}%
\bibitem [{\citenamefont {Segev}\ and\ \citenamefont {Van~de
  Walle}(2006)}]{Segev2006}%
  \BibitemOpen
  \bibfield  {author} {\bibinfo {author} {\bibfnamefont {D}~\bibnamefont
  {Segev}}\ and\ \bibinfo {author} {\bibfnamefont {C~G}\ \bibnamefont {Van~de
  Walle}},\ }\bibfield  {title} {\enquote {\bibinfo {title} {{Origins of
  Fermi-level pinning on GaN and InN polar and nonpolar surfaces}},}\
  }\href@noop {} {\bibfield  {journal} {\bibinfo  {journal} {EPL}\ }\textbf
  {\bibinfo {volume} {76}},\ \bibinfo {pages} {305--311} (\bibinfo {year}
  {2006})}\BibitemShut {NoStop}%
\bibitem [{\citenamefont {Ivanov}\ \emph {et~al.}(2010)\citenamefont {Ivanov},
  \citenamefont {Muljarov}, \citenamefont {Mouchliadis},\ and\ \citenamefont
  {Zimmermann}}]{ZimmermannComment2010}%
  \BibitemOpen
  \bibfield  {author} {\bibinfo {author} {\bibfnamefont {A.~L.}\ \bibnamefont
  {Ivanov}}, \bibinfo {author} {\bibfnamefont {E.~A.}\ \bibnamefont
  {Muljarov}}, \bibinfo {author} {\bibfnamefont {L.}~\bibnamefont
  {Mouchliadis}}, \ and\ \bibinfo {author} {\bibfnamefont {R.}~\bibnamefont
  {Zimmermann}},\ }\bibfield  {title} {\enquote {\bibinfo {title} {Comment on
  "photoluminescence ring formation in coupled quantum wells: Excitonic versus
  ambipolar diffusion"},}\ }\href {\doibase 10.1103/PhysRevLett.104.179701}
  {\bibfield  {journal} {\bibinfo  {journal} {Phys. Rev. Lett.}\ }\textbf
  {\bibinfo {volume} {104}},\ \bibinfo {pages} {179701} (\bibinfo {year}
  {2010})}\BibitemShut {NoStop}%
\end{thebibliography}%

\end{document}